\documentclass[11pt,a4paper]{article} 
\usepackage{jheppub}

\usepackage{slashed}
\usepackage{youngtab}

\newcommand\nn{\nonumber}

\newcommand{\beq}{\begin{equation}}
\newcommand{\eeq}{\end{equation}}

\newcommand{\beqa}{\begin{eqnarray}}
\newcommand{\eeqa}{\end{eqnarray}}

\newcommand{\bn}{\begin{equation}}
\newcommand{\en}{\end{equation}}
\newcommand{\by}{\begin{eqnarray}}
\newcommand{\ey}{\end{eqnarray}}
\newcommand{\al}{\alpha}
\newcommand{\be}{\beta}
\newcommand{\ga}{\gamma}
\newcommand{\Ga}{\Gamma}
\newcommand{\de}{\delta}
\newcommand{\ep}{\epsilon}
\newcommand{\om}{\omega}
\newcommand{\si}{\sigma}
\newfont{\bbbold}{msbm10 scaled \magstep1}
\def\a{\alpha}
\def\b{\beta}
\def\c{\gamma}
\def\d{\delta}
\def\ve{\varepsilon}
\def\f{\phi}\def\F{\Phi}
\def\h{\eta}

\def\k{\kappa}
\def\l{\lambda}\def\L{\Lambda}
\def\m{\mu}

\def\r{\rho}
\def\s{\sigma}\def\S{\Sigma}

\def\th{\theta}

\def\z{\zeta}

\def\o{\omega}\def\O{\Omega}
\def\bbR{\mbox{\bbbold R}}
\def\xz{\times}
\def\cA{{\cal A}}
\def\cB{{\cal B}}
\def\cC{{\cal C}}
\def\cD{{\cal D}}

\def\cI{{\cal I}}

\def\cM{{\cal M}}
\def\cN{{\cal N}}

\newcommand{\eq}[1]{(\ref{#1})}
\newcommand{\w}[1]{\\[0.#1cm]}

\def\tr{{\rm tr}}
\def\ua{\underline{\alpha}}
\def\ub{\underline{\phantom{\alpha}}\!\!\!\beta}
\def\uc{\underline{\phantom{\alpha}}\!\!\!\gamma}
\def\um{\underline{\mu}}

\def\half{\frac{1}{2}}

\title{Topologically gauged superconformal Chern-Simons matter theories}

\author[a]{Ulf~Gran,} 
\author[b,c]{Jesper~Greitz,}
 \author[c]{Paul~Howe}
 \author[a]{and Bengt~E.W.~Nilsson }
\affiliation[a]{Fundamental Physics\\ Chalmers University of Technology\\
SE-412 96 G\"oteborg, Sweden} 
\affiliation[b]{Nordita\\
Royal Institute of Technology and Stockholm University \\
Roslagstullsbacken 23, SE-106 91 Stockholm, Sweden }
\affiliation[c]{ Department of Mathematics, King's College London\\ The Strand, London WC2R 2LS, UK}
\emailAdd{ulf.gran@chalmers.se}\emailAdd{jesper.greitz@nordita.org} \emailAdd{paul.howe@kcl.ac.uk} \emailAdd{tfebn@chalmers.se}

\abstract{By coupling ${\cal N}=8$ superconformal  matter to ${\cal N}=8$ superconformal  Chern-Simons gravity in three dimensions we obtain theories with novel terms in the scalar potential leading to $AdS_3$ solutions and superconformal symmetry breaking. 
 If we start from the theory derived by Bagger, Lambert and Gustavsson, our coupled theory either inherits the $SO(4)$ gauge group or reduces it to $SO(3)$. If the construction is instead based on a free matter theory we find that the gravitational topological gauging also requires the introduction of a  Chern-Simons gauge sector   resulting in  a consistent theory for  any  $SO(N)$ gauge group. 
}


\begin{document}

\raisebox{0pt}[0pt][0pt]{\parbox{\textwidth}{
\begin{flushright}\small
KCL-MTH-12-03\\
NORDITA-2012-30
\end{flushright}}}

\maketitle  

\setcounter{page}{2}

\section{Introduction}

The study of theories living on stacks of branes in string/M-theory has proved to be a very fruitful area of research over the past few years. In this context the AdS/CFT correspondence  has led to crucial insights by providing a novel way to probe the non-perturbative aspects of these theories. While stacks of D-branes in string theory have turned out to be reasonably straightforward to model, the analogous case of stacks of M2 and M5-branes in M-theory have turned out to be much more difficult.  According to the AdS/CFT correspondence the physics of a stack of M2-branes should be captured, to leading order, by eleven dimensional supergravity in the near-horizon limit of the stack, i.e.~on the space $AdS_4 \times S^7$. In the dual picture, there should be a CFT describing the multiple M2-brane physics living on the boundary of $AdS_4$ which realises all the bulk symmetries.

A Chern-Simons (CS) matter theory \cite{Schwarz:2004yj} satisfying these requirements, i.e.~having superconformal ${\cal N}=8$ symmetry and $SO(8)$ R-symmetry, was found by Bagger and Lambert \cite{Bagger:2006sk,Bagger:2007jr} and Gustavsson \cite{Gustavsson:2007vu} (BLG). At the classical level the theory has a unique gauge group, $SO(4)=SU(2) \xz SU(2)$, which indicates an interpretation of  the theory as describing  two M2-branes. Subsequently, a lot of progress has been made by considering theories that have only ${\cal N}=6$ superconformal  symmetry manifestly realised \cite{Aharony:2008ug} then providing a description for stacks with any number branes. However,
this work and more recent results show that the ${\cal N}=8$ theory can be generalised at the quantum level to describe any number of M2-branes, see e.g.~\cite{Bashkirov:2011pt} and references therein. This also includes a better understanding of the $U(1)$ factor related to the centre of mass. 

An important ingredient in the AdS/CFT correspondence is the kind of boundary conditions that are imposed in the variational problem relating the bulk and boundary theories. The predominant boundary condition used is the Dirichlet one, where the variations on the boundary are required to vanish. This leads to CFTs in a fixed geometry, the BLG theory being one example of this. In this paper we will follow the ideas proposed in \cite{Nilsson:2012ky} and investigate the boundary theory that may be obtained by instead imposing Neumann or, more generally, mixed boundary conditions as discussed  in \cite{deHaro:2008gp}. 

In the Neumann variational problem, a consequence of the metric variations being non-vanishing at the boundary is that the boundary stress-energy tensor has to vanish. In order to allow for a non-vanishing stress-tensor we need to use mixed Dirichlet-Neumann boundary conditions, for which one has to add a gravitational Chern-Simons boundary term $\tfrac{k}{4\pi} S_{CS}$ to the bulk
 action\footnote{Note that while the bulk theory is parity 
 even the gravitational Chern-Simons boundary term, which enforces the mixed boundary conditions, is parity odd.}\cite{deHaro:2008gp}. For $k=0$ we recover the Neumann case, with vanishing stress-energy tensor, while $k=\infty$ sets the graviton fluctuations to zero at the boundary, thereby recovering the Dirichlet case. 
This and a number of other considerations relevant to this problem can be found in \cite{Nilsson:2012ky}.

In light of this it is natural to couple superconformal ${\cal N}=8$ matter, e.g. in the form of BLG theory, to (superconformal ${\cal N}=8$) Chern-Simons gravity
  in order to investigate the whole range of boundary conditions between Dirichlet and Neumann ones.  Such a coupled theory was first considered  in \cite{Gran:2008qx}
  where part of the theory was obtained. Using three different methods, {\it Noether, on-shell susy algebra} and {\it superspace},
   we will in this paper complete this computation and derive the most general theory of this type, 
 referred to as the topologically gauged $\cN=8$ Chern-Simons  matter theory (TGCS) in analogy with the $\cN=6$ case derived in \cite{Chu:2009gi} and further studied in
 \cite{Chu:2010fk}. One result of coupling superconformal ${\cal N}=8$ matter to superconformal ${\cal N}=8$ Chern-Simons gravity is that the  $S
 O(4)$ gauge group of the BLG theory is either left intact or reduced to $SO(3)$, but, more importantly, the coupling of conformal supergravity to the $free$ ${\cal N}=8$ matter  can be performed  for any gauge group  $SO(N)$. Thus, in this latter case the BLG gauge sector has been replaced by another one related to $SO(N)$. Furthermore, in this case there is
  only one Chern-Simons term as opposed to the two $SU(2)$ terms with opposite signs that occur in the BLG quiver model. If the gauged theory also has an interpretation in terms of branes these theories  would then describe  arbitrary stacks of  M2-branes, although it is not clear how many that would be for a given $N$.

The complete\footnote{Possibly modulo terms of cubic order in the Rarita-Schwinger field.} Lagrangian for superconformal ${\cal N}=8$ matter coupled to superconformal ${\cal N}=8$ Chern-Simons gravity is given in the next section, but focusing on the deformation of the BLG theory (i.e. $N=4$) we get, dropping terms with explicit gravitino fields,
\beqa
 L&=&\tfrac{1}{g}L_{conf}^{SUGRA}+L^{BLG}_{cov}-\tfrac{e}{16}X^2 R-V^{new}\notag\\[1mm]&&
+\tfrac{ig}{64}e(\bar\Psi_a\Psi_aX^2-10\bar\Psi_a\Psi_bX^I_aX^I_b+2\bar\Psi_a\Ga^{IJ}\Psi_bX^I_aX^J_b)
,~
\eeqa
where we see that the scalar potential, apart from the original contribution from the  $BLG$ theory, has received a new term given by
\begin{eqnarray}
V^{new} &=&   \tfrac{ eg^2}{2\cdot 32\cdot 32} \Big((X^2)^3 -8(X^2) X_b^J X_c^J X_c^K X_b^K +16X_c^I X_ a^I X_a^J X_b ^J X_b^K X_c^K \Big),
\end{eqnarray}
where $g$ is the (conformal) gravitational coupling constant \cite{Chu:2010fk} possibly related to a level parameter  of the gravitational Chern-Simons theory. The theory therefore depends on two parameters, the ordinary level parameter $\lambda=\tfrac{2\pi}{k}$, extracted as usual from the structure constants, and $g$. As we will see later these can be combined and interpreted as two ordinary level parameters related to the two CS terms in the quiver version of the theory. 
Note, however, that this new part of the potential is independent of the  structure constants $f^{abcd}$ and thus appears even if we start the gauging from the  free matter theory. In this case we therefore 
find an entirely new theory with gauge group $SO(N)$ that is not related to BLG at all and is consistent for any $N\geq 1$. 

In the above  Lagrangian  $L_{conf}^{SUGRA}$ is the action for superconformal ${\cal N}=8$ Chern-Simons gravity, see (\ref{LCSUGRA}) below, $L^{BLG}_{cov}$ is the BLG action (\ref{LBLG}) covariantised in the sense that the covariant derivatives also contain a spin connection and an $SO(8)$ R-symmetry gauge field. In addition one should note that
the usual BLG relation between $\tilde A_\mu{}^b{}_a$ (the 3-algebra connection) and $A_\mu{}^b{}_a$ (the field with respect to which the Lagrangian is varied) is generalised to 
\begin{eqnarray}
\tilde A_\mu^{ab} := A_\mu^{cd} (f_{cd}{}^{ab}-\tfrac{g}{4} \delta^{ab}_{cd})~.
\end{eqnarray}

The new term in the potential given above  is positive definite and  can be evaluated for a single or a multiple scalar vacuum expectation value (VEV) $v$. The resulting value of the potential  is of some importance when looking for background solutions. As will be discussed in more detail in section 2, this theory has  $AdS_3$ solutions corresponding to  topologically massive supergravity (TMSG) theories.
Interestingly enough, similar to the $\cN=6$ case where, as pointed out in \cite{Chu:2009gi,Chu:2010fk},  the $AdS$ solution is at a  chiral point in the sense of  \cite{Li:2008dq}, for the theory with $\cN=8$ derived here  the value $V(v)$ of the potential is different but does also seem to correspond to  critical values. In fact, depending on how many scalars $p$ are given the VEV $v$ one finds both critical $AdS$ and critical warped $AdS$ solutions \cite{Anninos:2008fx}. This difference between  ${\cal N}=8$ and ${\cal N}=6$ has of course consequences for how the spectrum is organised in the different cases
and should be studied further. We intend to return to this question elsewhere.

We also investigate whether the particular coupling to conformal supergravity we find in  $\cN=8$ has a counterpart for $\cN=6$ and find that it does not. This means that the results found in \cite{Chu:2009gi,Chu:2010fk} do not admit any further generalisation.

The paper is organised as follows: In section \ref{Bengt} the theory is derived using the Noether method in the 3-algebra formulation, thereby completing the results of \cite{Gran:2008qx}. In this section it is also explained how to compare  these results to the ones obtained in the algebraic and  superspace approaches presented in the following two sections. Thus in section \ref{Ulf} the theory is derived by closing the on-shell supersymmetry algebra generalising the method of the original references  \cite{Bagger:2006sk,Bagger:2007jr}. The derivation in superspace is given in section \ref{Paul}, and we end with  conclusions and a discussion in section \ref{Concl}. Some technical details can be found in the appendices.

\section{The Noether method}\label{Bengt}

The theory we  construct in this paper is the result
of turning the global symmetries of  $\cN=8$ superconformal matter  theories into local symmetries without destroying their conformal properties. The first example 
of this kind of gauging was given in  \cite{Gran:2008qx} where  it was applied to the BLG theory. The requirement of maintaining the conformal symmetries
implies that the supergravity sector  must itself consist of Chern-Simons (CS) terms, or  in other words, be topological. The topological aspects  will play an important r\^ole for what kind of degrees of freedom the gauged theories describe. That this kind of gauging does not add any new degrees of freedom (see e.g. \cite{Nilsson:2008ri}) was the main reason for the construction in  \cite{Gran:2008qx}. We emphasise again that the  construction also leads to a topologically gauged version of the free $\cN=8$ superconformal matter theory in which case we find a somewhat unexpected new result: The allowed gauge group is in this case $SO(N)$ for any positive integer $N$. Thus our results apply to stacks of branes as well as to a single brane provided the interpretation in terms of branes is still valid after the gauging. For a discussion of  this latter question, see  \cite{Nilsson:2012ky}. Since the gauged free matter  theories are unrelated to what is normally referred to as ``the BLG theory'' we will refer to all of them as just ``topologically gauged $\cN=8$ theories'' and use ``BLG''  only when  appropriate, i.e., when the gauge group is $SU(2)\times SU(2)$. Note that there is another $SO(4)$ theory with only one CS term, namely the theory obtained by gauging the free theory  for $N=4$. As will be clear later there is also a new theory which is a mixture of the two situations described above and that has only one $SO(3)$  as gauge group.

In this section we  give a brief but  hopefully  accessible account  of the topologically gauged $\cN=8$ theories and how to derive them using the Noether method.  The account of the complete theory given here  in terms of  the Lagrangian, transformation rules and field equations is written to facilitate the comparison to the results  of the
  other  two approaches. We also  hope it will benefit the reader to have the theory given  both in the three-algebra  and the quiver versions including an explanation of how to convert between them. Since the three different methods used in this paper to derive the same theory rely on different conventions we will  also
make an effort to relate them.

  We also take the opportunity to mention here that in the case of $\cN=6$ supersymmetries  we provide in a later section a new derivation in superspace proving   that 
 the topological gauging of \cite{Chu:2009gi} is unique. This can in fact also be seen in the Noether approach \cite{Chu:2009gi} since there one can analyse the cancellation options for the relevant terms  in $\de L$
 where this question arises. 

\subsection{The conformal supergravity sector}
We start by presenting  the pure conformal supergravity sector that is independent of the conventions used in the CS matter sector and is therefore  the same in the three-algebra and quiver  formulations.
 The fields in  on-shell $\cN=8$ conformal supergravity are the dreibein, $e_{\mu}{}^{\al}$, the Rarita-Schwinger field, $\chi_{\mu}$ with a hidden R-symmetry spinor index, and the R-symmetry gauge field, $B_{\mu}^i{}_j$.
 The topological CS Lagrangian is \cite{Deser:1982sw, VanNieuwenhuizen:1985ff, Lindstrom:1989eg, Gran:2008qx}
\begin{eqnarray}
L_{sugra}^{conf}&=&\tfrac{1}{2}\epsilon^{\mu\nu\rho}
Tr_{\alpha}(\tilde\omega_{\mu}\partial_{\nu}\tilde\omega_{\rho}+
\frac{2}{3}\tilde\omega_{\mu}\tilde\omega_{\nu}\tilde\omega_{\rho})
-i e^{-1} \epsilon^{\alpha\mu\nu}\epsilon^{\beta\rho\sigma}(\tilde
D_{\mu}\bar{\chi}_{\nu}\gamma_{\beta}\gamma_{\alpha}\tilde
D_{\rho}\chi_{\sigma})
\notag\\[1mm]
&&-\epsilon^{\mu\nu\rho}Tr_i
(B_{\mu}\partial_{\nu}B_{\rho}+\frac{2}{3}B_{\mu}B_{\nu}B_{\rho}),\label{LCSUGRA}
\end{eqnarray}
where the traces are over the three- and eight-dimensional vector indices  for the spin-connection and $SO(8)$ R-symmetry gauge field, respectively.
The tilde on the covariant derivatives refers to the spin connection  $\tilde\omega_{\mu \alpha\beta}$ which contains an ordinary second-order term  $\omega(e)$  plus a contorsion term $K$ bilinear in the Rarita-Schwinger field $\chi_{\mu}$. The covariant derivative is given explicitly in the subsection on the BLG theory below. Thus the three Chern-Simons terms in this Lagrangian are of third, second and first order in derivatives, respectively.

The standard procedure to obtain local supersymmetry for the spin-connection is to start by
adding Rarita-Schwinger terms to the dreibein-compatible $\omega(e)$. That is, one defines
\begin{equation}
\tilde\omega_{\mu \alpha\beta}=\omega_{\mu \alpha\beta}+K_{\mu
\alpha \beta},
\end{equation}
with
\begin{equation}
\omega_{\mu\alpha\beta}=\frac{1}{2}(\Omega_{\mu\alpha\beta}-\Omega_{\alpha\beta\mu}+\Omega_{\beta\mu\alpha}),\,\,\,\,
\Omega_{\mu\nu}{}^{\alpha}=\partial_{\mu}e_{\nu}{}^{\alpha}-\partial_{\nu}e_{\mu}{}^{\alpha},
\end{equation}
and
\begin{equation}
K_{\mu\alpha\beta}=-\frac{i}{2}(\chi_{\mu}\gamma_{\beta}\chi_{\alpha}-
\chi_{\mu}\gamma_{\alpha}\chi_{\beta}-\chi_{\alpha}\gamma_{\mu}\chi_{\beta}).
\end{equation}
This combination of spin connection and contorsion is
supercovariant, i.e. derivatives on the supersymmetry parameter
cancel out if $\tilde\omega_{\mu \alpha\beta}$ is varied under the
ordinary transformations of the dreibein and Rarita-Schwinger field:
\begin{equation}
\delta e_{\mu}{}^{\alpha}=i\bar\epsilon \gamma^{\alpha}\chi_{\mu},
\,\,\, \delta\chi_{\mu}= \tilde D_{\mu}\epsilon.
\end{equation}
The  above Lagrangian  is then supersymmetric if the R-symmetry gauge field varies according to
\beq
\de B_{\mu}^{ij}=-\frac{i}{2e}\bar \ep_g\Ga^{ij}\ga_ {\nu}\ga_{\mu}f^{\nu}+\de B_{\mu}^{ij}|_{new}
\eeq
where $f^{\mu}=\tfrac{1}{2}\ep^{\mu\nu\rho}\tilde D_{\nu}\chi_{\rho}$ and $\de B_{\mu}^{ij}|_{new}$ is a new term that is zero here but needed in the coupling to matter. With this latter  term present the end result
of the variation of the superconformal gravity sector is
\beq
\de L_{sugra}^{conf}=\ep^{\mu\nu\rho}\de B_{\mu}^{ij}|_{new}G^ {ij}_{\nu\rho}+\frac{i}{e}\bar f^{\mu}\ga_{\nu}\ga_{\mu}\Ga^{ij}\chi_{\rho}
\ep^{\nu\rho\si}\de B_{\si}^{ij}|_{new},
\eeq
where $G^{ij}_{\mu\nu}$ is the field strength of $B_{\mu}^{ij}$.

Following the strategy of \cite{Gran:2008qx}  in the Noether approach, we will use the first term in $\de L_{sugra}^{conf}$ above to determine $\de B_{\mu}^{ij}|_{new}$  from the cancellation of $\tilde D_{\mu}^2$ terms in the
variation of the Lagrangian. These new terms are all without derivatives  and will be  fed back into the computation through the second term in $\de L_{sugra}^{conf}$ above
which then gives rise to terms that are  first order in derivatives but quartic in fermions (counting also the susy parameter) and at least
bilinear in the $\chi_{\mu}$.

This supergravity theory is also invariant \cite{Gran:2008qx} under the following superconformal transformations with parameter $\eta(x)$:
\beq
\de_Se_{\mu}{}^{\al}=0,\,\,\,\de_S\chi_{\mu}=\ga_{\mu}\eta,\,\,\,\de_S  B_{\mu}^{ij}=\tfrac{i}{2}\bar\eta\Ga^{ij}\chi_{\mu},
\eeq
and dilatations with parameter $\phi(x)$:
\beq
\de_{\Delta}e_{\mu}{}^{\al}=-\phi e_{\mu}{}^{\al},\,\,\,\de_{\Delta}\chi_{\mu}=-\tfrac{1}{2}\phi\chi_{\mu},\,\,\,\de_{\Delta} B_{\mu}^{ij}=0.
\eeq
The strategy we will adopt in the Noether construction is to not allow $\delta\chi_{\mu}= \tilde D_{\mu}\epsilon$ to be extended by a term proportional to a $\ga$ matrix thereby eliminating the possibility of the S-supersymmetry to mix with ordinary supersymmetry in the derivation of the gauged theory.  Allowing for such a mixing  would just complicate the calculations without adding any extra information apart from proving also the superconformal invariance of the whole theory. However, this invariance is built into the superspace formulation and will be discussed more fully in that context.


\subsection{The covariantised matter sector in the 3-algebra formulation}


We now turn to the 3-algebra formulation of the CS-matter sector. It contains the fields of the $\cN=8$ BLG theory, i.e., eight real scalars, $X^i_a$ ,  in the vector representation
 (indices $i,j,k,..$) of the $SO(8)$ R-symmetry group   and eight Majorana spinors, $\psi_a$, in a chiral spinor representation of $SO(8)$ (here with non-explicit spinor  indices) together with a set of Chern-Simons gauge fields $A_{\mu a}{}^b$ in an as yet unspecified gauge group which acts on the 3-algebra index ($a,b,c,...$) possessed by both the scalar and spinor fields.
Recall that  the classical ungauged BLG theory in  flat space-time is consistent only for the gauge group $SO(4)$ while quantum mechanically the situation is rather  different (see, e.g., \cite{Bashkirov:2011pt}). The CS-matter Lagrangian explicitly covariantised under the new local symmetries diffeomorphisms, R-symmetry and supersymmetry reads \cite{Gran:2008qx}
\beqa
L_{BLG}&=&-\tfrac{1}{2}e\,g^{\mu\nu}\tilde D_{\mu}X^i_a\tilde D_{\nu}X^i_a+\tfrac{i}{2}e\bar\psi_a\ga^{\mu}\tilde D_{\mu}\psi_a
+\tfrac{1}{2}\epsilon^{\mu\nu\rho}
(A_{\mu ab}\partial_{\nu}\tilde A_{\rho}^{ab}+
\tfrac{2}{3}A_{\mu ab}\tilde A_{\nu}^{ac}\tilde A_{\rho}^{cb})\notag\\[1mm]
&&+\tfrac{i}{4}e\bar\psi_a\Ga^{ij}\psi_bX^i_cX^j_d\,f^{abcd}-\tfrac{1}{12}e(f^{abcd}X^i_bX^j_cX^k_d)(f^{aefg}X^i_eX^j_fX^k_g),\label{LBLG}
\eeqa
where we see that the scalar potential is positive definite and can be written as 
\beq
V(X)=\tfrac{\lambda^2}{12}(\ep^{abcd}X^i_bX^j_cX^k_d)(\ep^{aefg}X^i_eX^j_fX^k_g),
\eeq
after using the substitution of the structure constants by $\lambda\,\ep^{abcd}$ which is possible since the gauge group is $SO(4)$ in this case. Here $\lambda$ is related to the level $k$ by $\lambda=\tfrac{2\pi}{k}$.

The supergravity covariant derivative is
 \beq
 \tilde D_{\mu}\psi_a=\partial_{\mu}\psi_a+\tfrac{1}{4}\tilde \om_{\mu\al\be}\ga^{\al\be}\psi_a+\tfrac{1}{4}B_{\mu}^{ij}\Ga^{ij}\psi_a+\tilde A_{\mu ab}\psi_b,
 \eeq
 and
\beq
\tilde A_{\mu}^{ab}:=A_{\mu}^{cd}\,f^{cdab}.
\eeq
Here the structure constants are completely antisymmetric and indices are raised and lowered with a
delta (so we can be cavalier  about the position, up or down, of the three-algebra indices) to avoid problems with unitarity in the scalar field sector.  As a result of checking supersymmetry the structure constant must satisfy
the fundamental identity\footnote{The form of the identity given here was obtained in \cite{Gran:2008vi}. }
\beq
f^{[abc}{}_gf^{e]fg}{}_d=0.
\eeq
 In terms of finite dimensional 3-algebras, this identity is known to have only one solution corresponding to $\tilde A_{\mu}^{ab}:=A_{\mu}^{cd}\,f^{cdab}$ being an element of the  Lie algebra of $SO(4)$. The fact that $SO(4)$ is the unique gauge group is in general not changed by the gauging which can be seen in all three approaches as long as the structure constants are non-zero.
However, by tuning the two parameters it is possible to project away one of the two $SU(2)$ factors. Furthermore, it is very important to note  that after gauging one can set the ordinary BLG structure constants to zero, or just derive the gauged theory from a free $N=8$ matter theory. In either case one finds a theory without any  restrictions on the range of the 3-algebra indices and containing   only a single  Chern-Simons term with   a gauge group   $SO(N)$ for any $N$. This fact is clear in all three approaches.

The supersymmetry transformation rules are
\beq
\de X^i_a=i\bar\ep_m\Ga^i\psi_a,
\eeq
\beq
\de \psi_a=\ga^{\mu}\Ga^i\ep_m\tilde D_{\mu}X^i_a-\tfrac{1}{6}\Ga^{ijk}\ep_mX^i_bX^j_cX^k_d\,f^{abcd}+\de \psi_a|_{new},
\eeq
and
\beq
\de \tilde A_{\mu}^{ab}=i\bar\ep_m\ga^{\mu}\Ga^iX^i_c\psi_d\,f^{cdab}+\de \tilde A_{\mu}^{ab}|_{new},
\eeq
where we have indicated where the corrections will appear in the topologically gauged theory. Note that elimination of the BLG gauge fields by setting the structure constants to zero thus just means that a new set of  gauge fields are introduced as already mentioned.

Note also that the parameter $\ep_m$ has an index   $m$ for $matter$ since we will need to normalise it in a different way compared to the parameter in the supergravity sector
which we henceforth will denote as  $\ep_g$ in the Noether construction. This notation will also be useful when comparing the Noether approach to the algebraic and  superspace ones.


\subsection{The new interactions in gauged $BLG$  and the new $\cN=8$  $SO(N)$ theories}


 We now turn to the new interaction terms that arise as a result of the gauging and that are not already included in the covariantised version of the BLG theory discussed  above.  As in the ABJM case in \cite{Chu:2009gi} there are new interaction terms both with and without structure constants although some of the terms that appear in the $\cN=6$ case vanish identically in the $\cN=8$ situation discussed here. For example,
 using the antisymmetric structure constants above, it is easy to see that no sixth-order scalar potential term can be constructed with only one structure constant in the $\cN=8$ case. 
 
 There will appear  two parameters, $\lambda$ and $g$,  in the expressions below of which the former is related to the structure constant as
$f^{abcd}=\lambda\ep^{abcd}$, with $\lambda=\tfrac{2\pi}{k}$ and $k$ is the level.  The parameter $g$, on the other hand, is a dimensionless gravitational coupling constant
 (denoted $g^2_M$ in the analogues $\cN=6$ discussion  of  \cite{Chu:2010fk}) that must appear in such a way that the gravitational and matter sectors can be decoupled by
letting $g$ go to zero. While the way $g$ appears is the same in all three approaches, the way it is  introduced into the theory as well as its interpretation vary between the approaches. The two parameters will affect also the BLG part of the action and in particular the Chern-Simons term as will be explained in more detail below.

 Before doing that, however, we would like to emphasise that setting the structure constants, or $\lambda$ in the formulae below, to zero makes it possible to introduce a new  gauge group, namely $SO(N)$ for any $N$. This result is supported by the fact that, in any of three approaches, the range of the three-algebra index does not play any role in the calculation. Interestingly enough, there is a special new case with non-zero
$\lambda$ but with a gauge group that is different  from the BLG one, namely when the expression $\tfrac{g}{4}\de^{cd}_{ab}-\lambda \ep^{cdab}$ corresponds to
$P_{\pm}=\tfrac{1}{2}(\de\de\pm \tfrac{1}{2} \ep)$ which projects onto the self- and anti-selfdual parts. 
The  gauge group has then only one $SO(3)$ factor and hence only one CS term in the gauge sector.

 The Lagrangian derived here takes the following form, using $f^{abcd}=\lambda\ep^{abcd}$,
 \beqa
 L&&=\tfrac{1}{g}L_{conf}^{sugra}+L^{BLG}_{cov}
 + iAe\bar\chi_{\mu}\Ga^i\ga^ {\nu}\ga^{\mu}\psi_a(\tilde D_{\nu}X^i_a-\tfrac{i}{2}A\bar\chi_{\nu}\Ga^i\psi_a)
\notag\\[1mm]
&& -\tfrac{ie\lambda}{6}A\bar\chi_{\mu}\ga^{\mu}\Ga^{ijk}\psi_aX_b^iX_c^jX_d^k\,\ep^{abcd}+\tfrac{ie}{48}\lambda\bar\chi_{\mu}\ga^{\mu\nu}\Ga^{ijkl}\chi_{\nu}X_a^iX_b^jX_c^kX_d^l\,\ep^{abcd}
 \notag\\[1mm]&&
-\tfrac{i}{4}\ep^{\mu\nu\rho}\bar\chi_{\mu}\Ga^{ij}\chi_{\nu}(X^i_a\tilde D_{\rho}X^j_a)+iA\bar f^{\mu}\ga_{\mu}\Ga^i\psi_aX^i_a
+\tfrac{i}{4}X^2\bar f^{\mu}\chi_{\mu} -\tfrac{e}{16}X^2  R\notag\\[1mm]&&
+\tfrac{igA}{32}e\bar\chi_{\mu}\ga^{\mu}\Ga^i\psi_a(X^i_aX^2-4X^j_aX^j_bX^i_b)\notag\\[1mm]&&
-\tfrac{ig}{256}\ep^{\mu\nu\rho}\bar\chi_{\mu}\ga_{\nu}\chi_{\rho}((X^2)^2-4(X^i_aX^j_a)(X^i_bX^j_b))\notag\\[1mm]&&
+\tfrac{ig}{64}e(\bar\psi_a\psi_aX^2-10\bar\psi_a\psi_bX^i_aX^i_b+2\bar\psi_a\Ga^{ij}\psi_bX^i_aX^j_b)\notag\\[1mm]&&
-\tfrac{eg^2}{2\cdot 32\cdot 32}((X^2)^3-8(X^2)(X_a^iX_a^j)(X_b^iX_b^j)+16(X_a^iX_a^j)(X_b^jX_b^k)(X_c^kX_c^i)),
\eeqa
where $A^2=\tfrac{1}{2}$, $X^2:=X^i_aX^i_a$ and the covariant derivative is as given above. Note that the new potential appearing on 
last line in the Lagrangian can be written as a square as follows
\beq
V^{new}=\tfrac{eg^2}{2\cdot 32\cdot 32}((X^2)X^i_a-4(X_b^iX_b^k)X_a^k)^2.
\eeq

The full set of  supersymmetry transformation rules for the coupled theory is, with $\ep_m=A\ep_g$,
 \beq
\delta e_{\mu}{}^{\alpha}=i\bar\epsilon_g \gamma^{\alpha}\chi_{\mu},
\eeq
\beq
 \delta\chi_{\mu}= \tilde D_{\mu}\epsilon_g.
\eeq
\beqa
\de B_{\mu}^{ij}&=&-\tfrac{i}{2e}\bar \ep_g\Ga^{ij}\ga_ {\nu}\ga_{\mu}f^{\nu}-\tfrac{3ig}{8}\bar\psi_a\ga_{\mu}\Ga^{[i}\ep_mX^{j]}_a
-\tfrac{ig}{16}\bar\psi_a\ga_{\mu}\Ga^{ijk}\ep_mX^{k}_a\notag\\[1mm]
&&-\tfrac{ig}{4}\bar\chi_{\mu}\Ga^{k[i}\ep_gX^{j]}_aX^k_a-\tfrac{ig}{32}\bar\chi_{\mu}\Ga^{ij}\ep_gX^2
\eeqa
\beq
\de X^i_a=i\bar\ep_m\Ga^i\psi_a
\eeq
 \beqa
 \de \psi_a&=&\ga^{\mu}\Ga^i\ep_m(\tilde D_{\mu}X^i_a-iA\bar\chi_{\mu}\Ga^i\psi_a)-\tfrac{\lambda}{6}\Ga^{ijk}\ep_mX^i_bX^j_cX^k_d\,\ep_a{}^{bcd}\notag\\[1mm]
 &&+\tfrac{g}{8}\Ga^i\ep_mX^i_bX^j_bX^j_a-\tfrac{g}{32}\Ga^i\ep_mX^i_aX^2
   \eeqa
 \beqa
\de \tilde A_{\mu}^{ab}&=&-i\lambda\bar\ep_m\ga_{\mu}\Ga^i\psi_cX^i_d\,\ep^{cdab}-\tfrac{i\lambda}{2}\bar\chi_{\mu}\Ga^{ij}\ep_gX^i_cX^j_d\,\ep^{cdab}\notag\\[1mm]
&&+\tfrac{ig}{4}\bar\ep_m\ga_{\mu}\Ga^i\psi_{[a}X^i_{b]}+\tfrac{ig}{8}\bar\chi_{\mu}\Ga^{ij}\ep_gX^i_aX^j_b
\eeqa

We now give the main steps needed to derive this Lagrangian and the transformation rules in the Noether approach. The first step is to add the coupling term between the two sectors, the supergravity and CS-matter sectors, i.e., the $supercurrent$  term which is the  last term on the first line in the Lagrangian. It couples the Rarita-Schwinger field $\chi_{\mu}$ to the supercurrent  constructed to be conserved at the linear level in the fields\footnote{There is another conserved supercurrent which, however, can be seen to be equivalent to the one used here modulo terms involving the Dirac equation.}. In the standard fashion conservation of the supercurrent implies a gauge transformation of the gauge field, which 
in this case is just the supersymmetry transformation $\de\chi_{\mu}=\tilde D_{\mu}\ep_g$. 

The variation of the Lagrangian is then organised according to the number of covariant derivatives (counting field strengths as two derivatives) in each term. Cancelling all terms in $\de L$ of second order in derivatives (third-order terms occur only in the supergravity sector) requires the addition of new terms in the Lagrangian and in the transformation rules. These new terms are the ones  in the second and third lines in the Lagrangian (which are independent of the parameter $g$) and all the new terms in the variation of the gauge field $\de B_{\mu}^{ij}|_{new}$ while only the new $g$ independent term in  $\de \tilde A_{\mu}^{ab}|_{new}$ is derived at this stage in the calculation.  The term $\de\psi_a|_{new}$, on the other hand,  is determined entirely at order one derivative. This result was obtained already in \cite{Gran:2008qx} to which we refer the reader for the details of the calculation.

To cancel also the terms at order one derivative in $\de L$ we need to add the terms corresponding to $\de\psi_a|_{new}$ and the $g$ dependent terms in 
$\de \tilde A_{\mu}^{ab}|_{new}$. The remaining terms in the Lagrangian all arise at this level except the potential term which is multiplied  by $g^2$. At this point the  full set of transformation rules is determined.  The terms at this order that we need to prove cancel in $\de L$ are  with two spinors 
\beq
\ep D\psi X^3,\,\,\,\ep D\chi X^4,
\eeq
and with four spinors 
\beq
\ep D\chi \psi^2,\,\,\,\,\ep D\chi^2\psi X,\,\,\,\,\ep D\chi^3 X^2,
\eeq
where the last three have not been checked in detail.

Finally, checking that also the non-derivative terms cancel in $\de L$ will provide the coefficients of the new scalar potential terms proportional to $g^2$ but beyond that we only get   a number of cross-checks.  The terms involved here have two up to eight  fermionic quantities (fields together with the supersymmetry parameter) and  some of the cancellations will lead to  extensive fierzing. In the Noether approach, however,  we have only done the computations needed to obtain the final terms in the Lagrangian with one cross-check on the coefficients in the potential. 
The Lagrangian presented above is, however,  the complete answer possibly up to terms cubic in the Rarita-Schwinger fields. However, although we don't know for sure if terms with more than two Rarita-Schwinger fields can occur, we know from the Noether 
calculation with $\cN=6$ supersymmetries carried out  in \cite{Chu:2009gi} that some terms of this kind  were checked and seen not to appear. Modulo such cubic and higher terms in the Rarita-Schwinger field,  the only four-fermi term in the Lagrangian is the term that supercovariantises 
the the derivative in the supercurrent term. 

The terms without derivatives that have explicitly been checked to cancel are
\beq
\ep \chi X^6,\,\,\,\ep \psi X^3
\eeq
which fixes the coefficients in the potential and provides one cross-check on the result. In addition we have confirmed that the terms $\ep \psi^3 X$ cancel which requires some simple fierzing. This  calculation can be found in the Appendix.

We end this subsection with a discussion of the CS terms in the gauge sector since these have changed relative the covariantised BLG theory given above but 
we have not yet explained exactly what has happened. Looking at $\de \tilde A_{\mu}^{ab}$ it seems as if $f^{abcd}$ has been replaced by $f^{ab}{}_{cd}-\tfrac{g}{4}\de^{ab}_{cd}$  but this replacement  is not natural in other
sectors of the theory.  However, as can easily be checked, when discussing the CS term for $ A_{\mu}^{ab}$ this replacement is consistent with the requirements that one needs to put on 
 the variation of the CS term, for instance when obtaining the field equations. We will in fact  take advantage of this property of the CS term below.
That this is a correct procedure for dealing with the CS term  is supported by the results of all three approaches as will be demonstrated in the last  subsection.

Thus we end the discussion here by just quoting the answer:
\beq
L_{CS(A)}=\frac{1}{2}\epsilon^{\mu\nu\rho}
(A_{\mu ab}\partial_{\nu}\tilde A_{\rho}^{ab}+
\frac{2}{3} A_{\mu ab}\tilde A_{\nu}^{ac}\tilde A_{\rho}^{cb}),
\eeq
where, as  indicated by the form of $\de \tilde A_{\mu}^{ab}$, we have 
\beq
\tilde A_{\mu}^{ab}:=A_{\mu}^{cd}(\lambda \ep^{cdab}-\tfrac{g}{4}\de^{cd}_{ab}),
\eeq
which  is a direct generalisation of the ungauged BLG definition which corresponds to $g=0$.


\subsection{The field equations and $AdS_3$ background solutions}


We start with the field equations in the supergravity sector. The Cotton and Cottino equations are easily obtained from the Lagrangian given in the previous subsection  but since we will here primarily be interested in the background solution only the bosonic part of the Cotton equation is given.  

The variation with respect to the dreibein, or the metric if the spinors are set to zero, leads to the following Cotton equation
 \beqa
 &&\tfrac{1}{g}C_{\mu\nu}-\tfrac{eX^2}{16}(R_{\mu\nu}-\tfrac{1}{2}g_{\mu\nu}R)+
 \tfrac{e}{2}g_{\mu\nu}V(X)\notag\\[1mm]
 &&-\tfrac{e}{2}(D_{\mu}X^i_aD_{\nu}X^i_a-\tfrac{1}{2}g_{\mu\nu}D^{\si}X_a^iD_{\si}X_a^i)-\tfrac{e}{16}g_{\mu\nu}\Box X^2+\tfrac{e}{16}\nabla_{\mu}\nabla_{\nu}X^2=0.
 \eeqa
  The equation of motion for the R-symmetry gauge fields, on the other hand, will be useful to have in more detail. Up to $\chi$ dependent terms it reads
 \beq
\tfrac{1}{g} \ep^{\mu\nu\rho}G_{\nu\rho}^{ij}-eg^{\mu\nu}(\tilde D_{\nu}X_a^{[i})X_a^{j]}+\tfrac{i}{8}e\bar\psi_a\ga^{\mu}\Ga^{ij}\psi_a=0.
 \eeq

 Turning to  the matter sector we first give the scalar field equation. Discarding the fermions  it becomes
 $\Box X^i_a- \tfrac{1}{8}X^i_a R-\partial_{X^i_a}\,V(X)=0$
 which can be  seen to be  consistent with the trace of  the Cotton equation. In fact, combining these two scalar equations leads to the  condition on the potential
$X\partial_XV(X)=6V(X)$ which is obviously correct in a three-dimensional  conformally invariant theory. With the potential given above the full Klein-Gordon equation without $\chi$ dependent terms becomes
\beqa
&& \Box X^i_a- \tfrac{1}{8}X^i_a R-\tfrac{\lambda^2}{2}\ep_a{}^{bcg}X_b^jX_c^k\ep^{defg}X^i_dX^j_eX^k_f\notag\\[1mm]
&&-\tfrac{g^2}{32 \cdot 32}(3X^i_a(X^2)^2-8X_a^i(X_b^jX_b^k)(X_c^jX_c^k)-16X^2X_a^kX_b^kX_b^i+48X_a^j(X_b^jX_b^k)(X_c^kX_c^i))\notag\\[1mm]
&&-\tfrac{i\lambda}{2}\bar\psi_c\Ga^{ij}\psi_dX^j_b\ep_a{}^{bcd}+\tfrac{ig}{32}(\bar\psi_b\psi_bX^i_a-10\bar\psi_a\psi_bX^i_b+2\bar\psi_a\Ga^{ij}\psi_bX^j_b)=0.
\eeqa
The field equation for the gauge field $A_{\mu}^{ab}$  is, again discarding the $\chi$ terms,
\beq
\tfrac{1}{2} \ep^{\mu\nu\rho}\tilde F_{\nu\rho}^{ab}-eg^{\mu\nu}(\tilde D^{\mu}X^i_c)\tilde M_{cd}^{ab}X^i_d+\tfrac{i}{2}e\bar\psi_c\ga^{\mu}\psi_d\tilde M_{cd}^{ab}=0,
\eeq
where
\beq
\tilde M_{cd}^{ab}=\lambda \ep^{ab}{}_{cd}-\tfrac{g}{4}\de^{ab}_{cd}.
\eeq

The Dirac equation will also be useful in comparing  the results from the different approaches. It reads, discarding the $\chi$ dependent terms,
\beq
\ga^{\mu}\tilde D_{\mu}\psi_a+\tfrac{\lambda}{2}\Ga^{ij}\psi_bX^i_cX^j_d\ep^{cdab}+\tfrac{g}{32}(\psi_aX^2-10\psi_bX^i_aX^i_b+2\Ga^{ij}\psi_bX^i_aX^j_b)=0.
\eeq

We end this subsection by noting that similar to the $\cN=6$ case in  \cite{Chu:2009gi,Chu:2010fk} the above bosonic field equations are solved by a scalar vacuum expectation value (VEV) $v$  and an $AdS_3$ metric satisfying
\beq
R=-\tfrac{27}{128}v^4g^4.
\eeq
As  for $\cN=6$ we  find also in this case  that the vacuum solution corresponds to
a topologically massive supergravity. However,  there is an   important difference namely that while in $\cN=6$ the solution \cite{Chu:2009gi,Chu:2010fk} corresponds to a chiral point  in the sense of \cite{Li:2008dq} that is not quite the case here. Determining the relevant parameters as defined in \cite{Li:2008dq} we find for the theories discussed here that $\mu l=\frac{1}{3}$ while  the chiral point corresponds $\mu l=1$.
The value $\mu l=1$ is, however, obtained if we give two scalar fields the same VEV. Using the following form of the scalar VEV matrix\footnote{This   analysis may be carried out also for the new potential in the topologically gauged $\cN=6$ case \cite{Chu:2009gi}.}
\[ 
<X^i_a>=
 \left( \begin{array}{cc}
v{\bf 1_{p\times p}}& 0\\
0 & 0 
\end{array}
 \right),
  \]
 with $p=2$
 thus gives  an $AdS$ solution of the kind found in the $\cN=6$ case. Interestingly enough, by giving the same VEV to 3 or 6 scalars, i.e., for $p=3$ or $p=6$, we also find solutions  corresponding  to the critical value $\mu l=3$ in the analysis of warped $AdS_3$ in \cite{Anninos:2008fx}. Note   that ending up at some critical point is natural for reasons  having to do with massive graviton modes whose presence would be hard to explain from the point of view of the topological gauging, see for instance the discussion in \cite{Nilsson:2012ky}. This analysis still needs to be done in detail and will be discussed elsewhere. Finally, four scalars with the same VEV gives a vanishing $V^{new}$. Note that this discussion of solutions is valid for the case with zero BLG structure constants, i.e., for the new $SO(N)$ theories, and may be altered for $p\geq 3$ if the BLG structure constants are kept non-zero.

\subsection{Rewriting the results of  the Noether  3-algebra approach in quiver form}


We will now convert the theory above from the 3-algebra to the quiver formulation. The field equations written in this formulation  will then be easily compared  to the ones obtained
in the superspace approach which is also  in  quiver form.  As will be clear below, in the quiver formulation it is natural to introduce the two parameters discussed above as two level parameters tied to the
two chiral parts of the  gauge field since after the topological gauging they are no longer related to each other. However, we find that the gravitational Chern-Simons terms
in the gauged theory do not seem to make room for any additional level parameters although levels can be defined also in that sector,
see e.g.~Horne and Witten \cite{Horne:1988jf}. In fact, the three methods of derivation introduce the free parameters in different ways and seemingly for different reasons. Nevertheless, 
 they all point at the same result, namely that   the maximum number of free parameters (or levels) is just two.

\subsubsection{Conversion rules}

Starting from the real fields $X^i_a$ where $i$ takes 8 values and the 3-algebra index $a$ 4 values (as a vector of the  gauge group $SO(4)$) and introducing  
elements
$T^a$ of the three-algebra satisfying $Tr(T^aT^b)=\de^{ab}$  we define  $\hat X^i=X^i_aT^a$. Hence
\beq
L_{KG}=-\tfrac{1}{2}D^{\mu}X^i_aD_{\mu}X^i_a=-\tfrac{1}{2}Tr(D^{\mu}\hat X^iD_{\mu}\hat X^ i)=-tr(D^{\mu}X^iD_{\mu}X^{\dagger i}) ,
\eeq
where in the last step we have used
\[ 
\hat X=X_aT^a=
 \left( \begin{array}{cc}
0 & X\\
X^{\dagger} & 0 
\end{array}
 \right).
  \]

Note that the symbol $tr$ refers to the 2-dimensional trace over the indices of the matrices
\[ 
  X_m{}^{\bar m}=\frac{1}{2}(X^4{\bf 1} +i\si^iX^i)=\frac{1}{2}X^a\tau^a=\frac{1}{2}
 \left( \begin{array}{cc}
X^4+iX^3 & iX^1+X^2\\
iX^1-X^2 & X^4-iX^3 
\end{array}
 \right),
  \]
and
\[ 
  (X^{\dagger})_{\bar m}{}^{m}=\frac{1}{2}(X^4{\bf 1} -i\si^iX^i)=\frac{1}{2}X^a\bar \tau^a=\frac{1}{2}
 \left( \begin{array}{cc}
X^4-iX^3 & -iX^1-X^2\\
-iX^1+X^2 & X^4+iX^3 
\end{array}
 \right),
  \]
whose determinants are $SO(4)$ invariant.  

With the definitions above we have
 \[ 
T^a=\frac{1}{2}
 \left( \begin{array}{cc}
0 & \tau^a\\
\bar\tau^a & 0 
\end{array}
 \right),\,\,\,\, \tau^a=(i\si^i,{\bf 1}),\,\,\,\, \bar\tau^a=(-i\si^i,{\bf 1}).
  \]
 We can now define self-dual projections by considering 
   \[ 
T^{ab}=T^{[a}T^{b]}=\frac{1}{4}
 \left( \begin{array}{cc}
 \tau^{ab} & 0\\
0 & \bar\tau^{ab}  
\end{array}
 \right),
  \]
 and noting that  
$\tau^{ab}$  is self-dual and $\bar\tau^{ab}$  anti-selfdual.
  
  Turning to the triple product terms we have, from
  the definition of the three-algebra,
  \beq
  [T^a,T^b,T^c]=f^{abc}{}_dT^d,
  \eeq
  that 
  \beq
  X^i_aX^j_bX^k_cf^{abc}{}_dT^d=  X^i_aX^j_bX^k_c [T^a,T^b,T^c]=[\hat X^i,\hat X^j,\hat X^k].
  \eeq
  In terms of matrices this becomes
    \[ 
[\hat X^i,\hat X^j,\hat X^k]=4
 \left( \begin{array}{cc}
0 & (X^{[i}X^{\dagger j}X^{k]})_m{}^{\bar m}\\
(X^{\dagger [i}X^{ j}X^{\dagger k]})_{\bar m}{}^{ m} & 0 
\end{array}
 \right).
  \]
 
  With these definitions we can now translate the Lagrangian terms 
  \beqa
  L&=&-\tfrac{1}{2}D^{\mu}X^i_aD_{\mu}X^i_a-\tfrac{\lambda^2}{12}  X^i_aX^j_bX^k_cf^{abc}{}_d  X^i_{a'}X^j_{b'}X^k_{c'}f^{a'b'c'd}\notag\\[1mm]
  &&=-\tfrac{1}{2}Tr(D^{\mu}\hat X^iD_{\mu}\hat X^i)-\tfrac{\lambda^2}{12}Tr ([\hat X^i,\hat X^j,\hat X^k],[\hat X^i,\hat X^j,\hat X^k])
  \eeqa
  into the quiver formulation as \cite{VanRaamsdonk:2008ft}
  \beq
  L=-tr(DX^iDX^{\dagger i})-\tfrac{8\lambda^2}{3}tr((X^{[i}X^{\dagger j}X^{k]})(X^{\dagger [i}X^{ j}X^{\dagger k]})).
  \eeq
  Varying this with respect to $X^{\dagger i}$ we get, after collecting and rewriting the three terms that are obtained in the variation,
  \beq
  \Box X^i-8\lambda^2X^j(X^{\dagger [i}X^{ j}X^{\dagger k]}) X^k=0,
  \eeq
  where we also have inserted the level parameter $\lambda$ that can be  extracted from the structure constant. Comparing with the results in the superspace approach in section \ref{Paul} we find that
  \beq
 c^2=\lambda^2.
  \eeq
  Note that in this comparison the superspace expressions must be rescaled using $X\rightarrow \sqrt{2}X$ in order to get the supersymmetry transformations on the same form, which in fact also requires the susy parameter in superspace $\ep \rightarrow \sqrt{2}\ep:=\ep_m$, the Noether parameter in the matter sector.
  
  Repeating this for terms with four fields and one structure constant we find
  \beq
\tfrac{i\lambda}{4}\bar\psi_a\Ga^{ij}\psi_b X^i_cX^j_d\,\ep^{abcd}
=-2i\lambda tr(\bar\psi\Ga^{ij}\psi^{\dagger}X^iX^{\dagger j}),
  \eeq
  while terms without structure constants are simpler
  \beq
 \tfrac{i}{2}\bar\psi_a\ga^{\mu}\tilde D_{\mu}\psi_a +\tfrac{ig}{64}\bar\psi_a\psi_aX^2= itr(\bar\psi\ga^{\mu}\tilde D_{\mu}\psi^{\dagger})+\tfrac{ig}{16}tr(\bar\psi\psi^{\dagger})tr(X^iX^{\dagger i}).
  \eeq

When discussing the last type of expressions containing three fields that are not antisymmetrised, it is important to establish that the 3-algebra and quiver
formulations have a one-to-one relation. In particular we need to show that all quiver expressions in three fields can be written with one trace in order to translate them to the 3-algebra language. That this is indeed the case 
can be shown as follows. The set of five possible expressions with three quiver fields contains   two expressions with a trace
\beq
A=(X^{i})^{\bar m}{}_{m}tr(X^jX^{\dagger j}),\,\,B=(X^{j})^{\bar m}{}_{m}tr(X^iX^{\dagger j}),
\eeq
and three expressions without a trace
\beq
C=(X^iX^{\dagger j}X^j)^{\bar m}{}_{m},\,\,D=(X^jX^{\dagger i}X^j)^{\bar m}{}_{m},\,\,E=(X^jX^{\dagger j}X^i)^{\bar m}{}_{m},
\eeq
which, however, are linearly dependent. In fact, by cycling the two-dimensional indices it follows that  $A$ and $B$ may be used as a basis:
\beq
C=E=\tfrac{1}{2}A,\,\,\,D=B-\tfrac{1}{2}A.
\eeq

Next we use  the four-dimensional 3-algebra indices (assuming that the gauge group is $SO(4)$) to split the fields into self-dual and anti-self-dual parts. We consider first the covariant derivative 
\beqa
D_{\mu}X_a&&=\partial_{\mu}X_a+\tilde A_{\mu ab}X_b=\partial_{\mu}X_a+(2\lambda-\tfrac{g}{4}) A^+_{\mu ab}X_b-(2\lambda+\tfrac{g}{4}) A^-_{\mu ab}X_b\notag\\[1mm]
&&=\partial_{\mu}X_a+(2\lambda-\tfrac{g}{4})  A^+_{\mu ab}X_b+(2\lambda+\tfrac{g}{4})X_b A^-_{\mu ba}.
\eeqa
To get  these expressions into quiver form we multiply  by  3-algebra elements $T^a$ and  define, following \cite{VanRaamsdonk:2008ft}, the covariant derivative to be
\beq
D_{\mu}X_m{}^{\bar m}=\partial_{\mu}X_m{}^{\bar m}-(\tfrac{g}{8}-\lambda)  A^L_{\mu m}{}^{n}X_n{}^{\bar m}+
(\tfrac{g}{8}+\lambda)X_m{}^{\bar n} {A^R_{\mu\bar n}}{}^{\bar m}.
\eeq
To reach this form we have used the relations
\beq
A_{\mu}^{\pm}=\tfrac{1}{2}A_{\mu}^{L/R}=\tfrac{i}{2}A_{\mu}^{L/R4i}\si^i,
\eeq
where the gauge fields belong to the 
gauge group $SU_L(2)\times SU_R(2)$, i.e. the index structure is $A_{\mu}^{L4i}(\si^i)_m{}^n$ and $A_{\mu}^{R4i}(\si^i)_{\bar m}{}^{\bar n}$. 
As a final step we absorb the multiplicative factors into the gauge fields to get the covariant derivative in its standard form
\beq
D_{\mu}X_m{}^{\bar m}=\partial_{\mu}X_m{}^{\bar m}- A^L_{\mu m}{}^{n}X_n{}^{\bar m}+X_m{}^{\bar n} {A^R_{\mu\bar n}}{}^{\bar m}.
\eeq

Now we repeat these steps for the CS term where we should note the factor of $i$ in the definition of the quiver gauge fields above which therefore are anti-hermitian. Thus 
\beqa
\nn L_{CS(A)}&=&
\tfrac{1}{2}\epsilon^{\mu\nu\rho}
(2\lambda-\tfrac{g}{4})(A^+_{\mu ab}\partial_{\nu} A_{\rho}^{+ab}+
\tfrac{2}{3}(2\lambda-\tfrac{g}{8})A_{\mu ab}^+ A_{\nu}^{+ac}  A_{\rho}^{+cb})
\notag\\[1mm]&&
-\tfrac{1}{2}\epsilon^{\mu\nu\rho}
(2\lambda+\tfrac{g}{4})(A^-_{\mu ab}\partial_{\nu} A_{\rho}^{-ab}-
\tfrac{2}{3}(2\lambda+\tfrac{g}{4}) A_{\mu ab}^- A_{\nu}^{-ac}  A_{\rho}^{-cb}).
\eeqa
In terms of the (anti-hermitian) quiver fields this becomes
\beqa
\nn L_{CS(A)}&=&
\tfrac{1}{2}\epsilon^{\mu\nu\rho}
(\tfrac{g}{8}-\lambda)tr(A^L_{\mu }\partial_{\nu} A_{\rho}^{L}+
\tfrac{2}{3}(\tfrac{g}{8}-\lambda)A_{\mu}^L A_{\nu}^{L}  A_{\rho}^{L})
\notag\\[1mm]&&
+\tfrac{1}{2}\epsilon^{\mu\nu\rho}
(\tfrac{g}{8}+\lambda)tr(A^R_{\mu }\partial_{\nu} A_{\rho}^{R}+
\tfrac{2}{3}(\tfrac{g}{8}+\lambda) A_{\mu }^R A_{\nu}^{R}  A_{\rho}^{R}),
\eeqa
which after the same absorption of coefficients as above becomes
\beq
L_{CS(A)}=\tfrac{1}{a}L_{CS(A^L)}+\tfrac{1}{a'}L_{CS(A^R)},
\eeq
where we have identified the coefficients 
\beq
a:=\tfrac{g}{8}-\lambda,\,\,\,\,a':=\tfrac{g}{8}+\lambda.
\eeq
 These correspond in fact to  the ones used when setting up the superspace calculation, then defined
as the  levels for the two independent $SU(2)$ factors in the quiver formulation of the $SO(4)$ theory.

With these parameters present it is trivial to see that one can either turn off the BLG gauge group by letting $\lambda \rightarrow 0$ or the coupling to gravity by sending
$g \rightarrow 0$. In the latter case one must as usual first rescale away the inverse factor of $g$ in front of the gravitational Chern-Simons terms so that the limit
is not singular and all gravitational couplings to matter and gravitational self-interactions become  proportional to positive powers of $g$. This is done by sending the variation of the dreibein, the Rarita-Schwinger and R-symmetry fields to $\sqrt{g}$ times themselves. Note the gauge field $A_{\mu}^{ab}$ appears in a slightly different form  when using the above definition of the corresponding tilde field. To get an analogous situation for the gauge field it needs a redefinition by a square root of the  tilde factor but this can only be done after
splitting $A_{\mu}^{ab}$ into self-dual and anti-self-dual parts as done above.


\subsubsection{The   quiver formulation}

Here we use the conversion rules discussed above to present the result of rewriting the Lagrangian and transformation rules as found in the Noether approach in quiver form. Note that in doing so we continue to use the two supersymmetry parameters in the  Noether approach $\ep_g,\ep_m$ related by $\ep_g=2A\ep_m$, where $A^2=\tfrac{1}{2}$. 

We start with converting the matter sector. We find
\beqa
L_{BLG}&=&-e\,g^{\mu\nu}tr(\tilde D_{\mu}X^i\tilde D_{\nu}X^{\dagger i})+ie\,tr(\bar\psi\ga^{\mu}\tilde D_{\mu}\psi^{\dagger})\notag\\[1mm]
&&
+\tfrac{1}{2}a\epsilon^{\mu\nu\rho}
(A_{\mu}^L\partial_{\nu} A_{\rho}^{L}+
\tfrac{2}{3}aA_{\mu}^LA_{\nu}^{L} A_{\rho}^{L})+\tfrac{1}{2}a'\epsilon^{\mu\nu\rho}
(A_{\mu}^R\partial_{\nu} A_{\rho}^{R}+
\tfrac{2}{3}a'A_{\mu}^RA_{\nu}^{R} A_{\rho}^{R})\notag\\[1mm]
&&-2i\lambda\,e\,tr(\bar\psi\Ga^{ij}\psi^{\dagger}X^iX^{\dagger j})-\tfrac{8\lambda^2}{3}e\,tr((X^iX^{\dagger j}X^k)(X^{\dagger i}X^jX^{\dagger k})),\label{qLBLG}
\eeqa
where we used the definitions (and the notation $a,a',b,c$ from superspace) 
\beq
a=b+c=\tfrac{g}{8}-\lambda,\,\,\,a'=b-c=\tfrac{g}{8}+\lambda,
\eeq
The supergravity covariant derivative
 \beq
 \tilde D_{\mu}\psi=\partial_{\mu}\psi+\tfrac{1}{4}\tilde \om_{\mu\al\be}\ga^{\al\be}\psi+\tfrac{1}{4}B_{\mu}^{ij}\Ga^{ij}\psi-a\,A_{\mu}^L\psi+a'\,\psi A_{\mu}^R,
 \eeq
 and similarly for its complex conjugate $\psi^{\dagger}$.

The supersymmetry transformation rules are
\beq
\de X^i=i\bar\ep_m\Ga^i\psi,
\eeq
\beq
\de \psi=\ga^{\mu}\Ga^i\ep_m\tilde D_{\mu}X^i-\tfrac{2\lambda}{3}\Ga^{ijk}\ep_mX^iX^{\dagger j}X^k+\de \psi|_{new},
\eeq
and
\beq
\de  A_{\mu}^{L}=i\bar\ep_m\ga^{\mu}\Ga^i(X^i\psi^{\dagger}-\psi X^{\dagger i})+\de A_{\mu}^{L}|_{new},
\eeq
with a similar result for $A^R_{\mu}$ and where we have indicated where the corrections will appear in the topologically gauged theory.

Note that the parameter $\ep_m$ has an index   $m$ for $matter$ since we will need to normalise it in a different way compared to the parameter in the supergravity sector
which we henceforth will denote as  $\ep_g$ in the Noether construction. This will also be useful when comparing the Noether approach to the algebraic and  superspace ones.

We can then present the entire theory in the quiver formulation:
 \beqa
 L&&=\tfrac{1}{g}L_{conf}^{sugra}+L^{BLG}_{cov}
 + 2iAe\bar\chi_{\mu}\Ga^i\ga^ {\nu}\ga^{\mu}(tr(\psi\tilde D_{\nu}X^{\dagger i})-\tfrac{i}{2}Atr(\psi \bar\psi^{\dagger})\Ga^i\chi_{\nu})
\notag\\[1mm]
&& +\tfrac{4ie\lambda}{3}A\bar\chi_{\mu}\ga^{\mu}\Ga^{ijk}tr(\psi X^{\dagger i}X^jX^{\dagger k})-
\tfrac{ie\lambda}{6}\bar\chi_{\mu}\ga^{\mu\nu}\Ga^{ijkl}\chi_{\nu}tr(X^iX^{\dagger j}X^kX^{\dagger l})
 \notag\\[1mm]&&
-\tfrac{i}{2}\ep^{\mu\nu\rho}\bar\chi_{\mu}\Ga^{ij}\chi_{\nu}tr(X^i\tilde D_{\rho}X^{\dagger j})+2iA\bar f^{\mu}\ga_{\mu}\Ga^itr(\psi X^i)
+\tfrac{i}{2}tr(X^2)\bar f^{\mu}\chi_{\mu} -\tfrac{e}{8}tr(X^2)  R\notag\\[1mm]&&
+\tfrac{igA}{8}e\bar\chi_{\mu}\ga^{\mu}\Ga^i(tr(\psi X^i)tr(X^2)-4tr(\psi X^j)tr(X^{ij}))\notag\\[1mm]&&
-\tfrac{ig}{64}\ep^{\mu\nu\rho}\bar\chi_{\mu}\ga_{\nu}\chi_{\rho}((tr(X^2))^2-4tr(X^{ij})tr(X^{ij}))\notag\\[1mm]&&
+\tfrac{ieg}{16}(tr(\bar\psi\psi^{\dagger})tr(X^2)-10tr(\bar\psi X^{\dagger i})tr(\psi X^{\dagger i})+2tr(X^i\bar\psi^{\dagger})\Ga^{ij}tr(\psi X^{\dagger j}))\notag\\[1mm]&&
-\tfrac{eg^2}{16\cdot 16}((tr(X^2))^3-8tr(X^2)tr(X^{ij})tr(X^{ij})+16tr(X^{ij})tr(X^{jk})tr(X^{ki}))
\eeqa
where $A^2=\tfrac{1}{2}$ and we have introduced the short-hand notation $tr(X^{ij}):=tr(X^iX^{\dagger j})$ and $tr(X^2)=tr(X^{ii})$. The covariant derivative is as given above.

The full set of  supersymmetry transformation rules for the coupled theory is, with $\ep_m=A\ep_g$,
 \beq
\delta e_{\mu}{}^{\alpha}=i\bar\epsilon_g \gamma^{\alpha}\chi_{\mu},
\eeq
\beq
 \delta\chi_{\mu}= \tilde D_{\mu}\epsilon_g,
\eeq
\beqa
\de B_{\mu}^{ij}&=&-\tfrac{i}{2e}\bar \ep_g\Ga^{ij}\ga_ {\nu}\ga_{\mu}f^{\nu}-\tfrac{3ig}{4}tr(\bar\psi\ga_{\mu}\Ga^{[i}\ep_mX^{\dagger j]})
-\tfrac{ig}{8}tr(\bar\psi\ga_{\mu}\Ga^{ijk}\ep_mX^{\dagger k})\notag\\[1mm]
&&-\tfrac{ig}{2}\bar\chi_{\mu}\Ga^{k[i}\ep_gtr(X^{j]}X^k)-\tfrac{ig}{16}\bar\chi_{\mu}\Ga^{ij}\ep_gtr(X^2),
\eeqa
\beq
\de X^i=i\bar\ep_m\Ga^i\psi,
\eeq
 \beqa
 \de \psi&=&\ga^{\mu}\Ga^i\ep_m(\tilde D_{\mu}X^i-iA\bar\chi_{\mu}\Ga^i\psi)+\tfrac{2\lambda}{3}\Ga^{ijk}\ep_m(X^iX^{\dagger j}X^k)\notag\\[1mm]
 &&+\tfrac{g}{4}\Ga^i\ep_mtr(X^{ij})X^j-\tfrac{g}{16}\Ga^i\ep_mX^itr(X^2),
   \eeqa
 \beqa
\de A_{\mu}^L&=&-i\bar\ep_m\ga_{\mu}\Ga^i(\psi X^{\dagger i}\,-X^i\psi^{\dagger})-i\bar\chi_{\mu}\Ga^{ij}\ep_gX^iX^{\dagger j},\\
\de A_{\mu}^R&=&-i\bar\ep_m\ga_{\mu}\Ga^i(\psi^{\dagger}X^i-X^{\dagger i}\psi)\,-i\bar\chi_{\mu}\Ga^{ij}\ep_gX^{\dagger i}X^{ j}.
\eeqa





\subsection{Establishing the equivalence of the results in the different approaches}


In this subsection we establish the  relation between the three approaches and show that the three sets of results are equivalent.
This is most easily done by verifying that the final form of the transformation rules are the same which can be done after finding the proper relations between fields and parameters in three cases. There is a number of sign conventions that we will not try to sort out. Instead we will determine the connection between the parameters in the 
various approaches only up to signs but this will be enough to see that the most important features, like the form of the new potential terms, are exactly the same in all three approaches.

Consider first the transformation rules obtained in the Noether approach written in 3-algebra language and the ones derived in the algebraic  on-shell supersymmetry approach in the next section.
The equivalence between these two sets of transformation rules is established if one extracts $\lambda$ as usual from the structure constants in the latter case and set
\beq
\al^2=\tfrac{1}{16}g^2.
\eeq
In this case also signs can be made to match by noting the freedom we have to  place the gamma matrices in a prescribed order. This will produce signs in the algebraic approach  since there the gamma matrices   in the space-time directions anti-commute with those  in  the eight internal  directions.  

Next we recall the quiver form of the transformation rules in the Noether approach obtained in the previous subsection.
To translate the corresponding equations obtained in the superspace approach back to  the Noether approach we need to identify fields and parameters as follows, with $A^2=\tfrac{1}{2}$,
\beq
\Lambda=\psi,\,\,X_I=A^{-1}X^i,\,\,\,\ep=\ep_g=A^{-1}\ep_m,\,\,\,
\eeq
This will also establish the equivalence of the supersymmetry algebras  in the different approaches. Inspecting the transformation rules for the scalar and spinor fields in the matter sector we find that the 
equivalence between the Noether and  superspace approaches follows from the identifications:
\beq
b=\tfrac{1}{2}(a+a')=\tfrac{g}{8},\,\,\,c=\tfrac{1}{2}(a-a')=-\lambda,
\eeq
and hence
\beq
a=\tfrac{g}{8}-\lambda,\,\,\,a'=\tfrac{g}{8}+\lambda.
\eeq


\section{The SUSY algebra approach}\label{Ulf}


In this section we will outline how to derive the field equations of $\cN=8$ superconformal matter-coupled supergravity in three dimensions by requiring that the supersymmetry algebra closes. This method was used by Bagger and Lambert \cite{Bagger:2007jr} to derive the BLG theory and below we will see that extra structure is allowed, in the sense that the $SO(4)$ symmetry of the BLG theory can be replaced by $SO(N)$, by coupling to $\cN=8$ superconformal Chern-Simons gravity. The presentation here will focus on the method and results; for more technical details see appendix \ref{algebraapp}. The computations in this section have been facilitated by using the Mathematica package GAMMA \cite{Gran:2001yh}.


\subsection{Field equations of topologically gauged  $\cN=8$ BLG and $SO(N)$ theories}


Using the same conventions as in \cite{Gran:2008qx}, we have the following SUSY transformations for conformal Chern-Simons supergravity
\begin{eqnarray}
\delta e_{\mu}{}^{\alpha}&=& i\bar\epsilon \Gamma^{\alpha}\chi_{\mu}~,\notag \\
\delta\chi_{\mu}&=&\tilde D_{\mu}\epsilon~,\\
\delta B_{\mu}^{IJ}&=&-\frac{i}{2}e^{-1}\bar\epsilon\Gamma^{IJ}\Gamma_{\nu}\Gamma_{\mu}f^{\nu},\notag
\end{eqnarray}
where $\chi$ is the gravitino and $B_\mu^{IJ}$ is the $SO(8)$ gauge field. The supersymmetry transformations for conformal matter without coupling to gravity, as described by the BLG theory, are \cite{Bagger:2007jr}
\begin{eqnarray}
\delta X^{I}_a&=& i \bar \epsilon \Gamma^I \Psi_a~,\notag\\[1mm]
\delta\Psi_a&=&\bar D_{\mu}X^I_a \Gamma^{\mu}\Gamma^I \epsilon-\tfrac{1}{6}X^I_b X^J_c X^K_d \Gamma^{IJK}\epsilon f^{bcd}{}_a~,\\[1mm] 
\delta \tilde A_{\mu}{}^a{}_b &=& i\bar\epsilon \Gamma_{\mu} \Gamma^I X_c^I \Psi_d f^{cda}{}_b~.\notag
\end{eqnarray}
In order to couple these theories we make an ansatz for the SUSY transformations for the combined theory where we add all possible terms, to lowest order in the gravitino, allowed by the dimensions of the fields, arriving at\footnote{We have changed the sign of some of the standard BLG variations in order for the conformal Chern-Simons supergravity and BLG transformations to close into a translation having the same sign.} 
\begin{eqnarray}
\delta e_{\mu}{}^{\alpha}&=& \alpha_1 i\bar\epsilon \Gamma^{\alpha}\chi_{\mu}~, \notag\\
\delta\chi_{\mu}&=&\alpha_1\tilde D_{\mu}\epsilon +\alpha_2 \Gamma_\mu \epsilon X^2+ {\cal O}(\chi^2)~,\notag\\
\delta B_{\mu}^{IJ}&=&- \frac{i}{2}\alpha_1 e^{-1}(\bar\epsilon\Gamma^{IJ}\Gamma_{\nu}\Gamma_{\mu}f^{\nu})+i\alpha_3  (\bar \Psi_a \Gamma_\mu \Gamma^{[I}\epsilon) X^{J]}_a+i\alpha_4  (\bar \Psi_a \Gamma^K\Gamma^{IJ}\Gamma_\mu \epsilon) X^K_a\notag\\
&&+ i \alpha_8  (\bar\chi_\mu \Gamma^{K[I}\epsilon) X^{J]}_a X^K_a+ i \alpha_9 (\bar\chi_\mu \Gamma^{IJ} \epsilon) X^2 +{\cal O}(\chi^2),\notag\\
\delta X^{I}_a&=& i \bar \epsilon \Gamma^I \Psi_a~,\notag\\[1mm]
\delta\Psi_a&=&-(\bar D_{\mu}X^I_a-\frac{i}{\alpha_1}\bar \chi_\mu \Gamma^I \psi_a )\Gamma^{\mu}\Gamma^I \epsilon+\tfrac{1}{6}X^I_b X^J_c X^K_d \Gamma^{IJK}\epsilon f^{bcd}{}_a\\&&~+\alpha_5 \Gamma^I\epsilon X^I_a X^2+\alpha_6 \Gamma^I \epsilon X^J_a X^J_b X^I_b+ {\cal O}(\chi^2)~,\notag\\[1mm] 
\delta \tilde A_{\mu}{}^b{}_a &=&  i  \bar\epsilon \Gamma_{\mu} \Gamma^I X_c^I \Psi_d f^{cdb}{}_a- i\alpha_7  \bar \epsilon \Gamma_\mu \Gamma^IX^I_{[b}\Psi_{a]}+i \alpha_{10}  (\bar\epsilon \Gamma^{IJ}\chi_\mu) X^I_c X^J_d f^{cdb}{}_a\notag\\
&&+i\alpha_{11}  (\bar\epsilon \Gamma^{IJ}\chi_\mu) X^I_{b}X^J_a+ {\cal O}(\chi^2)~.\notag
\end{eqnarray}
where the constants $\alpha_i$ will be determined by requiring closure of the algebra. To simplify the ansatz we have used the freedom of making field redefinitions and the fact that two SUSY transformations of the fields are only allowed to contain $\partial \epsilon$-terms when acting on gauge fields\footnote{This in particular removes two possible $X$ times $\chi$ terms in $\delta X^I_a$.}.

The commutator of two supersymmetry transformations must close up to the symmetries of the theory; in our case general coordinate transformations, supersymmetry transformations, Lorentz transformations, $SO(8)$ and three-algebra gauge transformations and superconformal transformations. Note that all the global symmetries of BLG have been promoted to local ones when we couple to supergravity. 

Due to the simplicity of $\delta X^I_a$ we start by computing the commutator of two supersymmetry transformations on $X_a^I $ in order to identify the parameters of the symmetry transformations into which the SUSY algebra closes. Using the SUSY variations above we get
\begin{eqnarray}
[\delta_1, \delta_2] X_a^I &=& v^\mu \tilde D_\mu X^I_a + \delta_Q X^I_a- \Lambda^I{}_J X^J_a+\tilde\Lambda^b{}_a X^I_b~,
\end{eqnarray}
where
\begin{eqnarray}
v^\mu &=& 2 i \bar \epsilon_2 \Gamma^\mu \epsilon_1~,\notag\\\label{parameters}
Q &=& -\frac{1}{\alpha_1} v^\mu \chi_\mu~,\notag\\
\Lambda^{IJ} &=& -2 i  \alpha_5  (\bar \epsilon_2 \Gamma^{IJ} \epsilon_1) X^2 - 4 i \alpha_6 (\bar \epsilon_2 \Gamma^{[I|K}\epsilon_1)X^{J]}_b X^K_b)~,\\
\tilde\Lambda^b{}_a &=& i (\bar \epsilon_2 \Gamma^{KL} \epsilon_1) \big( X^K_c X^L_d f^{cdb}{}_a -2 \alpha_6 X^K_b X^L_a \big)~.\notag
\end{eqnarray}
We have here kept subleading terms in the gravitino in order to identity the supersymmetry parameter $Q$, but from now on we will only focus on the leading behaviour in the gravitino\footnote{This also means that we will not be careful regarding determinants of the vielbein in the expressions below as they will only give gravitino terms upon supersymmetry variation. The correct factors can be deduced when integrating the field equation to an action.}. 
In addition to the standard BLG terms \cite{Bagger:2007jr} there is now a local supersymmetry transformation, a local $SO(8)$ rotation and an additional term in $\tilde \Lambda^b{}_a$.

We now turn to the computation of $[ \delta_1,\delta_2] \Psi_a$. Using various Fierz rearrangements we find, to lowest order\footnote{There is no supersymmetry variation included in the RHS to lowest order in $\chi$ as $Q\sim \chi$.}  in $\chi$
\begin{eqnarray}
[\delta_1, \delta_2] \Psi_a &=& v^\mu \tilde D_\mu \Psi_a - \frac{1}{4}\Lambda_{IJ} \Gamma^{IJ}\Psi_a+\tilde\Lambda^b{}_a \Psi_b +\delta_S \Psi_a\notag\\
&&~+i(\bar \epsilon_2 \Gamma_\nu\epsilon_1)\Gamma^\nu \big(\cdots \big) \notag\\\label{ddpsi}
&&~-\frac{i}{4}(\bar \epsilon_2 \Gamma_{KL}\epsilon_1)\big(\cdots \big)\\
&&~+(\bar \epsilon_2 \Gamma_\mu \Gamma_{IJKL}\epsilon_1)\Gamma^\mu \big( \cdots \big)~,\notag
\end{eqnarray}
where $\delta_S$ denotes superconformal transformations and ellipses has been used to represent messy explicit formulae. To start, we note that we must have $\alpha_1 = \pm \sqrt{2}$ from the requirement that $[\delta_1,\delta_2] e_{\mu}{}^\alpha$ generates the same transport term as in $[\delta_1,\delta_2] X^I_ a$. 

We now analyse the different gamma-structures in (\ref{ddpsi}). The $\Gamma^{(5)}$ term does not represent symmetries of the theory, as can be seen from (\ref{parameters}), and therefore this term must vanish. From the $\Gamma^{(1)}$ term we can read off the Dirac equation, and finally the $\Gamma^{(2)}$ constitutes a consistency condition. Collecting the results\footnote{In additions four of the five parameters of the superconformal transformation are determined, see appendix \ref{algebraapp} for details.} we find
\begin{eqnarray}
\alpha_2 &=& \alpha_1 (2 \alpha_5 +\tfrac{1}{4}\alpha_7)~, \notag\\
\alpha_3 &=& 2\alpha_7~,\\
\alpha_6 &=& \tfrac{1}{2}\alpha_7~, \notag
\end{eqnarray}
and the Dirac equations is
\begin{eqnarray}
&&0=\slashed{\tilde D}\Psi_a +\tfrac{1}{2}\Gamma_{IJ}X^I_c X^J_d f^{cdb}{}_a \Psi_b +(3\alpha_5+\tfrac{\alpha_7}{4})X^2 \Psi_a - \tfrac{1}{\alpha_1} X^I_a \Gamma^I \Gamma^\mu f_\mu \notag \\
&&~~~~~~~+(3\alpha_4+2 \alpha_7)X^I_a X^I_b \Psi_b +(3\alpha_4+\tfrac{\alpha_7}{2})\Gamma_{IJ}X^I_a X^J_b \Psi_b~,\label{DE}
\end{eqnarray}
where we have included $\chi$-terms yielding curvatures upon variation as they will be important in the derivation of the Klein-Gordon equation below.

We now compute $[\delta_1,\delta_2]\tilde A_\mu{}^b{}_a$, which must take the form
\begin{eqnarray}
[\delta_1, \delta_2] \tilde A_\mu{}^b{}_a &=& -v^\nu \tilde F_{\mu\nu}{}^b{}_a + \tilde D_\mu \tilde\Lambda^b{}_a ~,
\end{eqnarray}
from which we can read off the field strength
\begin{eqnarray}
\tilde F_{\mu\nu}{}^b{}_a &=& \epsilon_{\mu\nu\lambda} \left( (X_c^I \tilde D^\lambda X_d^I -\tfrac{i}{2}\bar\Psi_c \Gamma^\lambda \Psi_d ) f^{cdb}{}_a  -\alpha_7 (X_{[b}^I \tilde D^\lambda X_{a]}^I -\tfrac{i}{2}\bar\Psi_{[b}\Gamma^{\lambda} \Psi_{a]})  \right)
\end{eqnarray}
and the constraints
\begin{eqnarray}
\alpha_{10} &=&  \tfrac{1}{\alpha_1}  ~,\notag\\
\alpha_{11} &=& -\tfrac{\alpha_7}{\alpha_1}~.
\end{eqnarray}
In this computation we need to use the fact that $f^{abcd}$ satisfies the fundamental identity. A further constraint from this computation is that a non-zero $f^{abcd}$ implies that the range of the three-algebra indices can be at most four. We thus see that in order to access the $SO(N)$ sequence of gauge groups for $N>4$ we need to set $f^{abcd}$, which parametrizes the BLG theory, to zero. Note however that when the range of the three-algebra indices is four we do get a deformation of the BLG theory, with the interesting self- and anti-self-dual cases discussed in section 2.

We now turn to closing the SUSY algebra on the gravitino. To lowest order in the gravitino it is only the superconformal transformation that contribute in the right-hand side of the algebra
\begin{eqnarray}
[\delta_1, \delta_2] \chi_\mu &=& \delta_S \chi_\mu+{\cal O}(\chi) ~.
\end{eqnarray}
The vanishing of terms not of this form implies that
\begin{eqnarray}
 \alpha_3+8\alpha_4 &=& 0~,
\end{eqnarray} 
and the requirement of matching the superconformal transformation read off from closing the supersymmetry algebra on $\Psi_a$ yields
\begin{eqnarray}
\alpha_5 &=& -\tfrac{1}{8} \alpha_7~.
\end{eqnarray}

Let us now consider the $SO(8)$ gauge field. From the requirement that 
\begin{eqnarray}
[\delta_1, \delta_2] B_\mu{}^{IJ} &=& -v^\nu G_{\mu\nu}{}^{IJ} + \tilde D_\mu \Lambda^{IJ} ~,
\end{eqnarray}
we can read off the $SO(8)$ field strength
\begin{eqnarray}
G_{\mu\nu}{}^{IJ} &=& 2 \alpha_7 \epsilon_{\mu\nu}{}^\rho ( X_a^{[I} \tilde D_\rho X_a^{J]}-\tfrac{i}{8}\bar \Psi_a \Gamma_\rho \Gamma^{IJ} \Psi_a)\label{SO8FS}
\end{eqnarray}
and the constraints
\begin{eqnarray}
\alpha_8 &=& 4\tfrac{\alpha_6}{\alpha_1} ~,\notag \\
\alpha_9 &=& -2 \tfrac{\alpha_5}{\alpha_1}~. 
\end{eqnarray}
The constraints obtained on the $\alpha_i$ parameters leave only one free parameter, which we choose to be $\alpha_7$ and relabel as just $\alpha$ from now on.

Making a supersymmetry variation of the Dirac equation (\ref{DE}) which, using the solution for the $\alpha_i$ parameters, simplifies to
\begin{eqnarray}
&&0=\slashed{\tilde D}\Psi_a +\tfrac{1}{2}\Gamma_{IJ}X^I_c X^J_d f^{cdb}{}_a \Psi_b -\tfrac{\alpha}{8}X^2 \Psi_a +\tfrac{5}{4}\alpha X^I_a X^I_b \Psi_b  \notag \\
&&~~~~~~~ -\tfrac{1}{4}\alpha\Gamma_{IJ}X^I_a X^J_b \Psi_b - \tfrac{1}{\alpha_1} X^I_a \Gamma^I \Gamma^\mu f_\mu~,
\end{eqnarray}
we obtain the Klein-Gordon equation
\begin{eqnarray}
\tilde \Box X_a^I-d_{X_a^I} V -\tfrac{1}{8}X^I_a \tilde R &=& 0 \label{KG}
\end{eqnarray}
where
\begin{eqnarray}
V &=& \tfrac{1}{12} X_a^I X_b^J X_c^K X_e^I X_f^J X_g^K f^{abcd} f^{efg}{}_d + \tfrac{\alpha^2}{8} \Big( \tfrac{1}{16}(X^2)^3 \notag\\
&&~~-\tfrac{1}{2}(X^2) X_b^J X_c^J X_c^K X_b^K +X_c^I X_ a^I X_a^J X_b ^J X_b^K X_c^K \Big) ~.
\end{eqnarray}
By instead making a supersymmetry variation of the $SO(8)$ field strength (\ref{SO8FS}) we get the Cottino equation
\begin{eqnarray}
&& 4 e^{-1} \epsilon_\sigma{}^{\mu\nu}\Gamma_\rho \Gamma_\mu \tilde D_\nu f^\rho  +\tfrac{\alpha^2}{\alpha_1} \Gamma_\sigma\Gamma^I \Psi_b (X_b^K X_a^K X_a^I -\tfrac{1}{4}X_b^I X^2)\notag \\
&& +2\tfrac{\alpha}{\alpha_1}\Gamma^I (\Psi_a \tilde D_\sigma X_a^I - \tilde D_\sigma\Psi_a  X_a^I) +2\tfrac{\alpha}{\alpha_1} \Gamma^\rho \Gamma_\sigma \Gamma^I \Psi_a \tilde D_\rho X_a^I \\
&& - \tfrac{1}{3} \tfrac{\alpha}{\alpha_1} \Gamma_\sigma \Gamma^{IJK} X_a^I X_b^J X_c^K \Psi_d f^{abcd} + \alpha e^{-1} \Gamma_\rho \Gamma_\sigma f^\rho X^2= 0\notag ~,
\end{eqnarray}
which can be rewritten using the Dirac equation giving
\begin{eqnarray}
&& 4 e^{-1} \epsilon_\sigma{}^{\mu\nu}\Gamma_\rho \Gamma_\mu \tilde D_\nu f^\rho - 2\tfrac{\alpha}{\alpha_1} \epsilon_\sigma{}^{\mu\nu} \Gamma_\mu \Gamma^I \tilde D_\nu (\Psi_a X^I_a) +4\tfrac{\alpha}{\alpha_1} \Gamma^\rho \Gamma_\sigma \Gamma^I \Psi_a \tilde D_\rho X_a^I +2 \alpha e^{-1}  f_\sigma X^2 \notag\\
&&  -2\tfrac{\alpha^2}{\alpha_1} \Gamma_\sigma\Gamma^I \Psi_b (X_b^K X_a^K X_a^I -\tfrac{1}{4}X_b^I X^2)  + \tfrac{2}{3} \tfrac{\alpha}{\alpha_1} \Gamma_\sigma \Gamma^{IJK} X_a^I X_b^J X_c^K \Psi_d f^{abcd} = 0 ~. \label{Cottino}
\end{eqnarray}
Performing a supersymmetry variation of the Cottino equation (\ref{Cottino}) yields the Cotton equation
\begin{eqnarray}
&-\tfrac{4}{\alpha}\tilde C_{\mu\nu} + X^2 (\tilde R_{\mu \nu} -\tfrac{1}{2}g_{\mu\nu}\tilde R) -8 g_{\mu\nu} V(X) + g_{\mu\nu} \tilde \Box X^2 - \tilde D_{(\mu} \tilde D_{\nu)} X^2\notag\\
&~~~~~~+8(\tilde D_\mu X_a^I \tilde D_\nu X_a^I-\tfrac{1}{2}g_{\mu\nu}\tilde D^\rho X_a^I \tilde D_\rho X_a^I) =0
\end{eqnarray}
where we have only kept the purely bosonic terms.

And finally, by using the Klein-Gordon equation (\ref{KG}) and the fundamental identity we can show that the Bianchi identity
\begin{eqnarray}
\epsilon^{\mu\nu\rho} \tilde D_\mu \tilde F_{\nu\rho}{}^a{}_b &=& 0
\end{eqnarray}
is satisfied. 

To summarise, starting from a general ansatz for the supersymmetry transformations we have been able to derive the field equations for conformal matter coupled to conformal Chern-Simons gravity in three dimensions  by requiring that the supersymmetry algebra closes up to symmetry transformations of the theory. The most notable outcome of coupling to gravity is that the allowed gauge group is $SO(N)$ for $N\geq 1$.



\section{Superspace}\label{Paul}


In this part of the paper we discuss these models from the superspace point of view. The fields can be thought of as belonging to three multiplets: conformal supergravity, Yang-Mills and scalar. The first of these can be described off-shell for any $\cN$, and as this has been discussed previously in the literature \cite{Howe:1995zm,Howe:2004ib,Kuzenko:2011xg,Greitz:2011vh,Cederwall:2011pu} we have relegated it to appendix D. However,  the treatment given there is quite detailed and contains some previously unpublished material on the completeness of the solution to the Bianchi identities and $\cN=6$ as well as the conventions for this section. The basic feature of the off-shell conformal supergravity field strength multiplet, for $\cN\geq 4$, is that it is given by a constrained dimension-one Lorentz-scalar superfield, $M_{IJKL}$, that is totally antisymmetric on its $SO(\cN)$ indices. This superfield contains the Cotton and Cottino tensors amongst its components and can therefore be thought of as the super Cotton tensor. A similar situation holds in the Yang-Mills sector where there is another constrained dimension-one superfield $W_{IJ}$ \cite{Samtleben:2009ts,Samtleben:2010eu}. The constraints are given in \eq{2.13} for gravity and \eq{2.262} for Yang-Mills. For conformal supergravity this leads to the multiplet structure given in Figure 1 in appendix D, while a similar picture, given by Figure 2, is valid for the Yang-Mills sector. On the other hand the scalar multiplets for $\cN=6$ and $\cN=8$ are on-shell. They both consist of  (sets of) eight scalars and spinors so the basic constraints, which state that the former transform into the latter, necessarily imply the equations of motion by supersymmetry. 

To construct the complete on-shell theory we therefore need to specify the super Cotton tensor $M_{IJKL}$ and the Yang-Mills tensor $W_{IJ}$ in terms of the scalar multiplet fields. Since these both have dimension one it follows that they must be bilinears in the scalar fields, and from the basic constraint on the latter it follows that the constraints on $M$ and $W$ are satisfied, and hence that the Bianchi identities for both gravity and Yang-Mills are also satisfied, as discussed in D.2.

The problem is therefore to check that these equations are consistent with the Ricci identity for the scalar multiplet.\footnote{By Ricci identity we mean the definition of a curvature in terms of a graded commutator of derivatives.} For the latter, the anti-commutator of two odd superspace derivatives acting on the scalar, or the spinor, will give rise to terms involving the torsion, the geometrical curvatures and the Yang-Mills field strength, and it turns out to be sufficient to check this anti-commutator for the scalars. From this consistency constraint, given that the Bianchi identities in the gauge and geometry sectors hold,  the restrictions on the possible gauge groups that are allowed can be derived. By applying further odd derivatives one can then obtain the equations of motion for the spacetime fields.

In the following we shall occasionally refer to $(p,q)$ forms, superspace forms with $p$ even and $q$ odd indices. For example, the Bianchi identity for a Yang-Mills field is $\cI:=DF=0$, and the lowest-dimensional component of this is the one with all odd indices, i.e. $\cI_{0,3}$. 


\subsection{Matter multiplets}


For a vector field in $D=3$ the conformal Lagrangian is the Chern-Simons term so that in the absence of any matter the equation of motion states that the spacetime field strength is zero, and the superspace extension of this is simply that the whole of the superspace field strength vanishes. When matter is present the spacetime field strength will be given by the dual of the matter current and in superspace  this implies that all components of the field strength will be given as bilinears in the matter fields. There have been discussions of these models in the literature in a superspace context in \cite{Cederwall:2008vd,Cederwall:2008xu,Bandos:2008jv}, from a pure spinor inspired point of view, in \cite{Bandos:2008df,Bandos:2009dt} using superfields and the Nambu bracket formulation of BLG, and in \cite{Samtleben:2009ts,Samtleben:2010eu} in a more conventional approach that is close to ours.

 In this section we shall briefly recap what happens in flat superspace in a Lie-algebra formalism. We shall take the gauge group to be $G \xz G'$ in $\cN=6$ and recover the result that $G=G'=SU(2)$ for $\cN=8$.

We begin with $\cN=8$.\footnote{For $\cN=8$ the R-symmetry group is taken to be $Spin(8)$ rather than $SO(8)$ whence the spinor R-symmetry index $A$. See appendix D.} The scalar multiplet has eight scalars $X_I$ and eight spinors $\L_{\a A'}$, so the constraint on the superfield $X_I$ must be

\beq
D_{\a A} X_I =i(\S_I)_{AA'} \L_\a^{A'}\ .
\label{3.1}
\eeq
Here, the derivative is gauge-covariant with respect to the group $G \xz G'$, so the Ricci identity is

\beq
[D_{\a A},D_{\b B}] X_I=i\d_{AB}(\c^a)_{\a\b} D_a X_I - aF_{\a A\b B} X_I +a' X_I F'_{\a A\b B}\ ,
\label{3.2}
\eeq
where $a,a'$ are real constants.
To check the consistency of the Ricci identity we can parametrise the variation of $\L$ as

\beq
D_{\a A} \L_{\b B'}=\half (\c^a)_{\a\b} (\S_I)_{AB'} D_a X_I + \ve_{\a\b} H_{AB'} + (\c^a)_{\a\b} H_{a AB'}\ .
\label{3.3}
\eeq
The first term is there in the absence of interactions while the fields appearing the second and third will be functions of the matter fields as we are on-shell.  In order to determine these we need first to say something about the lowest-dimensional components of the fields strengths. The scalar $X$ has dimension one-half, $\L$ dimension one and $F_{0,2}$ also has dimension one. The latter can therefore only be bilinear in $X$. We have\footnote{In other words these equations give $W_{AB}$ and $W'_{AB}$ in terms of the scalars.}

\beq
F_{\a A\b B}=i \ve_{\a\b}(\S^{IJ})_{AB} X_I X^*_J\qquad {\rm and }\qquad F'_{\a A\b B}=i\ve_{\a\b} (\S^{IJ})_{AB} X^*_I X_J\ ,
\label{3.4}
\eeq
Using \eq{3.1} we can easily see that these constraints are compatible with the Bianchi identities for $F,F'$. The dimension three-halves component of $F$ is given by

\beq
F_{a\b B}=(\c_a\chi)_{\b B}\ ,
\label{3.5}
\eeq
and similarly for $F'$ where 

\beqa
\chi_{\a A}&=& i (\S^I(X_I\L^* - \L X^*_I))_{\a A}\nn\w1
\chi'_{\a A}&=& i (\S^I(X^*_I\L - \L^* X_I))_{\a A}\ .
\label{3.6}
\eeqa
As discussed in appendix D it suffices to show that the $(0,3)$ component of the Bianchi identity holds, which it clearly does
given equations \eq{3.1} and \eq{3.4}

Substituting \eq{3.3} into \eq{3.2} we find, in the case of flat superspace, that $H_{aBC'}=0$, and that

\beq
2(\S_I)_{[A}{}^{C'} H_{B] C'}=(\S^{JK})_{AB}\left(b(X_JX^*_KX_I-X_IX^*_JX_K)+c(X_JX^*_KX_I+X_IX^*_JX_K)\right)\ ,
\label{3.7}
\eeq
where $b:=\half(a+a')$ and $c:=\half(a-a')$. The terms cubic in $X$ contain the representations $(1000)$, $(0011)$ and $(1100)$ of $SO(8)$, but only the first two are contained in $H_{AB'}$.  The mixed symmetry representation must therefore be excluded. It appears inevitably in the $b$ term, so that $b$ must be set to zero, i.e. $a=-a'$. This means that the coefficients of the two Chern-Simons terms in the spacetime action must have equal magnitude and opposite sign. The $c$ term will also have a mixed symmetry component except for the case $SU(2)\xz SU(2)$, and when $X_I$ is real,

\beq
X_x{}^{x'} \rightarrow \bar X^x{}_{x'}=\ve^{xy} X_y{}^{y'}\ve_{y'x'}\ \forall\ I\ ,
\label{3.8}
\eeq
where $x,x'=1,2$ are doublet indices for the two $SU(2)$s. In this case, it is easy to see, using the cyclic formula

\beq
A B^* C+ AC^* B=A \tr (B^*C)\ ,
\label{3.9}
\eeq
valid for any real fields (as in \eq{3.8}) in the bi-fundamental representation of $SU(2)\xz SU(2)$. In this case one finds that only the totally antisymmetric $X^3$ term survives and that

\beq
H_{AA'}=\frac{c}{6} (\S^{IJK})_{AA'} X^3_{IJK}\ ,
\label{3.10}
\eeq
where $X^3_{IJK}:=X_{[I} X^*_J X_{K]}\ .$

In the ABJM case the scalar field $Z_A$ is complex, in the four-dimensional spinor representation of $SU(4)$, the spin group of $SO(6)$. As it is complex it can also carry a $U(1)$ charge $q$ with respect to the additional $U(1)$ R-symmetry factor.  In flat superspace the basic equations are: the variation of the scalar,

\beq
D_{\a I} Z_A =i(\S_I)_{AB} \L_\a^{B}\ .
\label{3.11}
\eeq
the Ricci identity,

\beq
[D_{\a I},D_{\b J}] Z_A=i\d_{IJ}(\c^a)_{\a\b} D_a Z_A - aF_{\a I\b J} Z_A + a' Z_A F'_{\a I\b J}\ ,
\label{3.12}
\eeq
the variation of $\L$,

\beq
D_{\a I} \L_{\b}^B=\half (\c^a)_{\a\b} (\S_I)^{BC} D_a Z_C + \ve_{\a\b} H_I{}^B + (\c^a)_{\a\b} H_{a I}{}^B\ ,
\label{3.13}
\eeq
and the dimension-one components of the gauge field strengths,

\beq
F_{\a I\b J}=ia \ve_{\a\b}\,Z\S^{IJ}Z^* \qquad {\rm and }\qquad F'_{\a I\b J}=ia'\ve_{\a\b}\, Z^*\S^{IJ}Z\ .
\label{3.14}
\eeq

As in the the BLG case we can substitute \eq{3.13} in \eq{3.12} to find, firstly, that $H_{a I}{}^B=0$, and then that

\beq
4\S_{[I} H_{J]}=b(Z\S^{IJ}Z^*Z-ZZ^*\S^{IJ}Z) + c(Z\S^{IJ}Z^*Z+ZZ^*\S^{IJ}Z)\ .
\label{3.15}
\eeq
Making the $SU(4)$ structure of these terms more explicit we find that the $b$ term is

\beq
2b(\S_{IJ})^B{}_C Z_{(A}\bar Z^C Z_{B)}\ ,
\label{3.16}
\eeq
whereas the $c$ term is antisymmetric on $AB$. \eq{3.16} can be rewritten as 

\beq
2b(\S^{KLM}\S_{IJ})_{AB} \xi_{KLM}{}^B\ .
\label{3.17}
\eeq
The spinor $\xi$ can be expanded in terms of irreducible representations and we find that it contains  the thirty-six dimensional  $(201)$ representation that does not drop out of \eq{3.15} and that cannot be absorbed in $H_I$. So we again have to choose $b=0$. On the other hand, the $c$ term is compatible with \eq{3.15} for any choice of gauge group of the form $G \xz G'$. One finds

\beq
H_I{}^A=c(\z_I-\frac{1}{4}\S_I \S^J\z_J)^A\ ,
\label{3.18}
\eeq
where

\beq
\z_I{}^A:=-\half (\S_I)^{BC} Z_B\bar Z^A Z_C\ .
\label{3.19}
\eeq


\subsection{Coupling to supergravity}


In this section we consider the coupling of the matter-gauge systems to conformal supergravity. The idea is that we have to satisfy the Bianchi identities in the gravity and gauge sectors and the Ricci identity for the matter fields. For the ABJM case it turns out that the parameter $b$ must still be set equal to zero and that the scalar multiplet can be coupled to the off-shell superconformal geometry, while for the BLG case the situation is more complicated. There one can couple the scalar multiplet to on-shell conformal supergravity, but only if the parameter $b$ is non-zero.


\subsubsection{$\cN$=6}


The basic constraint on the scalar multiplet \eq{3.11} is unchanged (although the derivative now includes the geometrical connections), and the dimension-one components of the gauge field strength tensors are also unaltered. However, the Ricci identity \eq{3.12} is amended to 

\beq
[D_{\a I},D_{\b J}] Z_A=i\d_{IJ}(\c^a)_{\a\b} D_a Z_A - aF_{\a I\b J} Z_A +a'  Z_A F'_{\a I\b J}-iq G_{\a I\b J}Z_A-R_{\a I\b J,A}{}^B Z_B\ ,
\label{4.1}
\eeq
where $q$ is the $U(1)$ charge of the scalar field $Z$ and the last term involves the $SO(6)$ curvature in the spin representation, $R_A{}^B=\frac{1}{4}(\S^{IJ})_A{}^B R_{IJ}$. The variation of $\L$ is still given by \eq{3.13}, although the field $H_a$ cannot be set to zero. We now want to investigate the Ricci identity on the scalar fields using the variation of the spinor, as before. The crucial terms come from the $M\xz Z$ terms in the $SO(6)$ and $U(1)$ curvatures acting on $Z$. The key point is that these give rise to a composite object $M_{IJ} Z_A$, (where, for $\cN=6$, $M_{IJ}$ is the $SO(6)$ dual of the super Cotton tensor $M_{IJKL}$), which decomposes into $(201)+(011)+(100)$. The matter multiplet, as we discussed above, does contribute a term in the $(201)$ when $b\neq 0$, as can be seen from \eq{3.16}. However, $M_{IJ}$ can only be proportional to $\tr(Z\S_{IJ} Z^*)$ on-shell, whereas \eq{3.16} cannot be written in this form for any choice of gauge group $G \xz G'$, not even for $G=G'=SU(2)$. In the $SU(2)\xz SU(2)$ case one can use \eq{3.9} to write \eq{3.16} as a sum of two terms involving traces. One of these has the correct form to be absorbed by $M_{IJ} Z_A$, but the other has the form $tr (ZZ) Z^*$ which cannot. We therefore conclude that $b=0$ in the presence of conformal supergravity as well as in flat space.

This result means that the $(201)$ representation in $M_{IJ} Z$ cannot be absorbed by the matter sector and hence the two $MZ$ terms in \eq{4.1} must be arranged so that this term cancels between them. This requires the charge $q$ to be $-\half$. With this choice the susy algebra on the scalars closes provided that the $H$-functions in the variation of $\L$ are chose to be

\beqa
H_{a I}{}^A&=&L_{aIJ}(\S^J Z)^A -\half L_a{}^{JK}(\S_{IJK} Z)^A\nn\w1
H_I{}^A&=&\frac{1}{2} K_{IJ}(\S^J Z)^A+c(\z_I-\frac{1}{4}\S_I \S^J\z_J)^A-i\m_I{}^A +\frac{3i}{8} (\S_I\S^J\m_J)^A \ ,
\label{4.2}
\eeqa
where

\beq
M_{IJ} Z:=\hat \m_{IJ} +\S_{[I}\m_{j]}\ .
\label{4.3}
\eeq
Here, $\hat\m_{IJ}$ denotes the $(201)$, while $\m_I$ is the sum of the other two representations, $(011)$ and $(100)$. Note that this does not require that the conformal supergravity sector is on-shell, although of course one will need to impose this to obtain the full equations of motion. This is done by  taking $M_{IJ}$ to be proportional to $\tr(Z\S_{IJ} Z^*)$. The conclusion is therefore that the $\cN=6$ gauging of \cite{Chu:2009gi,Chu:2010fk} does not admit any further generalisation.


\subsubsection{$\cN$=8}


The situation is somewhat different in $\cN=8$. We shall start with the closure of supersymmetry on the scalars, for which \eq{3.2} is modified to

\beq
[D_{\a A},D_{\b B}] X_I=i\d_{AB}(\c^a)_{\a\b} D_a X_I - a F_{\a A\b B} X_I + a' X_I F'_{\a A\b B}-R_{\a A\b B,I}{}^J X_J\ ,
\label{4.4}
\eeq
while the variation of the fermion is given in \eq{3.3}. We now find that there is a solution for the $H$-fields in the variation of $\L$ for the gauge group $SU(2)\xz SU(2)$ with non-vanishing conformal supergravity provided that $b\neq 0$. Explicitly we find

\beqa
H_{AA'}&=&\frac{1}{2}(\S^I)_{A'}{}^B K_{AB} X_I +\frac{b}{2}(\S^I)_{AA'}(\tr(X_IX^*_J)X_J-\frac{1}{4} \tr X^2 X_I)\nn\w1
&\phantom{=}&-\frac{c}{6}X^3_{IJK}\nn\w1
H_{aAA'}&=&\frac{1}{4}(-(\S^I)_{AA'} L_{a IJ} X_J+\frac{1}{2}(\S^{IJK})_{AA'} L_{a IJ} X_K)\ ,
\label{4.5}
\eeqa
where $\tr X^2:=\tr( X_I X^*_I)$. The $b$ term in $H_{AA'}$ can be absorbed by the geometry provided that we choose 

\beq
C_{IJ}=8b(\tr(X_I X^*_J)-\frac{1}{8} \d_{IJ} \tr X^2)\ ,
\label{4.6}
\eeq
where $C_{IJ}$ is the super Cotton tensor for $\cN=8$. (Its relation to $M_{ABCD}$ is described in Appendix D.)
It might be thought that the terms in \eq{4.5} involving $K_{AB}$ and $L_{a BC}$ could be ignored since their leading components can be gauged away (see Appendix D), but this is not correct because the spinorial derivatives of these fields include terms involving the gravitino field strength and the  field $\l_{ABC}$ (the dimension three-halves component of the super Cotton multiplet). In particular, the latter turns out to be (using \eq{2.26})

\beq
\l_I=-\frac{14ib}{3}(\tr (X_I\L^*)-\frac{1}{7}\S_{IJ} \tr (X_J \L^*))\ .
\label{4.7}
\eeq

This essentially solves the problem in superspace. It is tedious, but straightforward, to verify that the supersymmetry algebra closes on $\L$ and to obtain as a bi-product the equation of motion for the spinor field. It is

\beqa
\c^aD_a\L&=&-\frac{b}{4}(2\S^{IJ}\mathrm{tr}(\L X_I)X_J+10\mathrm{tr}(\L X_I)X_I-\mathrm{tr}X^2\L)\nn\w1 
									&\phantom{=}&+ c\S^{IJ}X_I\L X_J -
									\frac{3}{4}\S^I\Psi X_I+\frac{1}{4}L_{aIJ}\c^a\S^{IJ}\L\label{4.8}
									\eeqa
From this one finds the pure scalar terms in the scalar equation of motion to be

\beqa
D^a D_aX_I&=& b^2 ( 3\tr(X_IX_J) \tr(X_J X_ K)X_K-\tr X^2 \tr(X_I X_J)X_J-\half \tr(X_J X_K)\tr(X_J X_K)X_I+\nn\w1
&\phantom{=}&+ \frac{3}{16}(\tr X^2)^2X_I) +2c^2X_JX^3_{IJK}X_K\ .
\label{4.10}
\eeqa
The last term can be rewritten in terms of traces if desired. This corresponds to a potential that is proportional to

\beqa
V(X)&\propto& b^2\left( \half \tr(X_I X_J)\,\tr(X_I X_K)\,\tr(X_J X_K)   -\frac{1}{4}\tr X^2\,   \tr(X_I X_J)\,\tr(X_I X_J)   +\frac{1}{48}(\tr X^2 )^3\right)\nn\w1
&\phantom{\propto}&-c^2\tr( X^3_{IJK}X^3_{IJK})\ .
\label{4.11}
\eeqa

For the geometrical sector the general analysis given in appendix D shows that we have a complete solution of the Bianchi identities provided that the field $M_{ABCD}$ satisfies \eq{2.13} (with the indices replaced by $A,B$ etc and with $\l_5$ being the dual of $\l_3$). That this is so is easily verified from \eq{4.6} and \eq{3.1}. For the gauge sector we have already shown that the Bianchi identities are satisfied given \eq{3.4} and \eq{3.1}. The dimension three-halves components are given by \eq{3.5} and \eq{3.6}.  The dimension-two components are given by

\beqa
F_{ab}&=&-\ve_{ab}{}^c(X_I D_c X^*_I-D_c X_I X^*_I+\frac{i}{2}\L_{A'}\c_c\L^*_{ A'}) \nn\w1
F'_{ab}&=&-\ve_{ab}{}^c(X^*_I D_c X_I-D_c X^*_I X_I+\frac{i}{2}\L^{*}_{A'}\c_c\L_{A'})\ .
\label{4.12}
\eeqa

The equations derived above are covariant under super-Weyl transformations, although this is not manifest. A discussion of this topic can be found in Appendix D.


\subsection{$\cN=8$ models with $SO(N)$ gauge groups}


In the $\cN=8$ modification of BLG we have seen that setting $b=0$ immediately implies that the background has to be superconformally flat, because $C_{IJ}$ and hence all of the field strengths in the super Cotton tensor must vanish. On the other hand one can set $c=0$ without getting a free model. If the gauge group is $SU(2)\xz SU(2)$ this means that the two Chern-Simons terms in the spacetime Lagrangian have equal magnitudes and signs. They can therefore be rewritten as a single $SO(4)$ Chern-Simons term. It turns out, as we shall now show, that this model can be generalised to an $SO(N)$ gauge group for any $N$.

We now take the scalar field $X_I^r,r=1,\ldots N$ to transform under the vector representation of $SO(N)$ as well as $SO(8)$. Then \eq{4.4} becomes

\beq
[D_{\a A},D_{\b B}] X^r_I=i\d_{AB}(\c^a)_{\a\b} D_a X_I - aF^{rs}_{\a A\b B} X^s_I -R_{\a A\b B,I}{}^J X^r_J\ ,
\label{4.23}
\eeq
while in \eq{3.2} both $X$ and $\L$ carry an extra $SO(n)$ vector index and \eq{3.4} is replaced by

\beq
F_{\a A\b B}^{rs}=i(\S^{IJ})_{AB} X^r_I X^s_J\ .
\label{4.24}
\eeq
We then find that the supersymmetry algebra closes on the scalars if we take

\beq
C_{IJ}=8a(\tr(X_I X_J)-\frac{1}{8}\d_{IJ} \tr X^2)\ 
\label{4.25}
\eeq
and

\beq
H_{AA'}=\frac{1}{2}(\S^I)_{A'}{}^B K_{AB} X_I+ \frac{a}{2}(\S^I)_{AA'}(\tr (X_I X_J) -\frac{1}{4}\d_{IJ} \tr X^2)X_J\ .
\label{4.26}
\eeq


\section{Conclusions}\label{Concl}


 The main goal of this paper  has been  to derive the possible topologically gauged M2-brane theories with six and eight supersymmetries. We have done this by
 combining results from three different methods, partly from expediency and partly because the different approaches illuminate different aspects of the problem with the results appearing in a different orders.
Clearly each of these methods, {\it Noether}, {\it algebraic on-shell supersymmetry}  and {\it superspace},  would  by itself
 provide a complete and consistent  way to derive any of the  topologically gauged  theories discussed in this paper. 
 
In the case of $\cN=6$ supersymmetry we have shown rigorously that there are no further possibilities for topological gauging of the ABJM models beyond those that have already appeared in the literature \cite{Chu:2009gi,Chu:2010fk}. On the other hand, for $\cN=8$, we have seen that the gauged BLG theory can accommodate an extra interaction in the presence of gravity provided that the coefficients of the two $SU(2)$ Chern-Simons terms no longer have the same magnitude. When the standard BLG sextic potential is switched off this results in a new interacting theory with a single $SO(4)$ Chern-Simons term. Furthermore, this model can be extended to arbitrary $SO(N)$ gauge groups. Note, however, that for special values of the parameters in the deformed  BLG the $SO(4)$ is reduced to $SO(3)$.

One general feature of these topologically gauged theories is the appearance of new terms in the scalar potential. 
If evaluated for a single scalar field vacuum expectation value $v$, the new potential picks up 
 a non-zero value $V(v)$  which  implies that all these theories have $AdS_3$ vacuum solutions of various kinds. In the $\cN=6$ cases considered in \cite{Chu:2009gi,Chu:2010fk} the resulting topologically massive supergravity theory ended up at the chiral point corresponding to that discussed by Li, Song and Strominger in \cite{Li:2008dq}. However,  the  theories  with eight supersymmetries obtained  in this paper have different solutions depending on how many scalar fields are given the VEV $v$. As explained in section 2,  in this way we find both critical $AdS_3$ backgrounds and  critical warped  $AdS_3$ ones corresponding to those discussed in \cite{Anninos:2008fx}. 

The interpretation of these new models in terms of membranes is not immediately apparent. We have argued in the introduction that they might correspond to branes in the context of 
AdS/CFT with mixed Dirichlet-Neumann boundary conditions  as suggested in \cite{Nilsson:2012ky}. It would be interesting to study whether this tentative identification can be established with more certainty.

\acknowledgments
We would like to thank Martin Cederwall and Jelle Hartong for discussions and Nordita and the organisers of the programme ``Geometry of strings and fields" where part of this work was carried out. BN is partly funded by the Swedish Research Council and UG is supported by the Knut and Alice Wallenberg Foundation.

\appendix

\section{Comparing conventions}

\subsection{Conventions in the Noether  approach}
We use here the convention
\beq
\ep^{\mu\nu\rho}:\,\,\,\,\ep^{012}=+1
\eeq
together with
\beq
e\ga^{\mu\nu}=\ep^{\mu\nu\rho}\ga_{\rho},\,\,\,\,
\tfrac{1}{2}\ep^{\mu\nu\rho}\ga_{\nu\rho}=-e\ga^{\mu},\,\,\,
\ep^{\mu\nu\rho}\ep_{\tau\nu\rho}=-2e^2\de^{\mu}_{\tau}\,.
\eeq

The eight internal gamma matrices $\Ga^i$ are $16\times 16$, antisymmetric and commute with the ones in space-time $\ga^{\mu}$.

\subsection{Supersymmetry conventions}

The choice of the dimension zero torsion in superspace $T_{\al\be}{}^a=-i\ga^a_{\al\be}$ implies using the superspace Ricci identity (the curvature term is not needed here)
\beq
\{D_{\al},D_{\be}\}=-T_{\al\be}{}^a D_a=i\ga^a_{\al\be}D_a\,,
\eeq
which can be turned into a commutation relation for two supersymmetries as follows. Form $\de_1=\bar\ep_1^{\al}D_{\al}$ and the same for the second one, and 
construct the commutator. We find that
\beq
[\de_1,\de_2]=i\bar\ep_1\ga^a\ep_2D_a\,.
\eeq

Now we compute the same commutator from the transformation rules of the component fields in the Noether and algebraic approaches. Using
\beq
\de X^i_a=i\bar\ep\Ga^i\psi_a,\,\,\,\de\psi_a=\Ga^i\ga^{\mu}\ep D_{\mu}X^i_a\,,
\eeq
(note the order of the gamma matrices in $\de \psi_a$ which makes it valid in both component approaches) we find that
\beq
[\de_1,\de_2]=-2i\bar\ep_1\ga^{\mu}\ep_2D_{\mu}\,.
\eeq
The difference in the factor of "2" is clear since it is taken care of by the factor of $\tfrac{1}{2}$ in the superspace approach, $A$ in the Noether approach and the 
$\sqrt{2}$ in the algebraic approach. The sign difference on the other hand is correct since the derivative and field realizations should give susy algebras with different signs on the right hand side.

\subsection{ $(\ga^{\mu}, \Ga^i)$ versus $\Ga^M=(\Ga^{\mu}, \Ga^I)$  versus $(\ga^a,(\Sigma^I)_A{}^{A'})$}

In the Noether and superspace approaches the 3d gamma matrices commute with the 8d ones, while in the algebraic approach they anti-commute.
These choices are not significant and could have been done differently in each case. In the title of this subsection the matrices used in the three approaches are 
given in the order Noether, algebraic, and superspace.  There is also some notational differences for the $SO(8)$ gamma matrices. In the component approaches one uses $\Ga^i$, with $i$ an $SO(8)$ vector index, which in superspace is replaced by $\Sigma^I$ and $\bar\Sigma^I$, now with $I$ as the vector index. One can check that the Clifford algebra in this case allows for two different kinds of gamma matrices. While the metric in spinor space must be symmetric the $\Ga^i$ matrices themselves can be either symmetric or antisymmetric. The Noether  approach uses antisymmetric $\Ga^i$ matrices but in the superspace approach this is not an issue since only the $8\times 8$ blocks
$\Sigma^I$ and $\bar\Sigma^I$ appear. If needed 
the relation between the gamma matrices  is
\[
\Gamma^i=
 \left( \begin{array}{cc}
0 & \Sigma^I \\
\bar\Sigma^I & 0
\end{array}
 \right)\,,
 \]
 where $\bar\Sigma^I=- \Sigma^I $ while in superspace it is understood that if needed $\bar\Sigma^I= \Sigma^I $ is used.
Note that the chirality of the R-symmetry spinors is explicit in  the superspace conventions where the gamma matrices  are often written with spinor indices as $(\Sigma^I)_A{}^{A'}$.

 \section{Cancellation of  $\ep \psi^3 X$ in the Noether approach}
 
One interesting term to check here is $\ep \psi^3 X$ without a structure constant. Such terms  come from the following variations
\beq
\de L_{Dirac}|_{\de B}=\tfrac{ie}{8}\bar\psi_a\ga^{\mu}\Ga^{ij}\psi_a\de B_{\mu}^{ij},
\eeq
\beq
 \de L_{Dirac}|_{\de A, no f}=\tfrac{ie}{2}\bar\psi_a\ga^{\mu}\psi_b\de \tilde A_{\mu}^{ab},
 \eeq
 and 
 \beq
\de L_{Yuk, no f}|_{\de X}=\tfrac{ig}{32}e(\bar\psi_a\psi_aX_b^k\de X^k_b-10\bar\psi_a\psi_bX^i_a\de X^i_b+2\bar\psi_a\Ga^{ij}\psi_bX^i_a\de X^j_b).
\eeq
 Inserting the relevant variations we get
 \beq
\de L_{Dirac}|_{\de B}=\tfrac{ie}{8}\bar\psi_a\ga^{\mu}\Ga^{ij}\psi_a( -\tfrac{3ig}{8}\bar\psi_b\ga_{\mu}\Ga^{[i}\ep_mX^{j]}_b
-\tfrac{ig}{16}\bar\psi_b\ga_{\mu}\Ga^{ijk}\ep_mX^{k}_b),
\eeq
\beq
 \de L_{Dirac}|_{\de A, no f}=\tfrac{ie}{2}\bar\psi_a\ga^{\mu}\psi_b\tfrac{ig}{4}\bar\ep_m\ga_{\mu}\Ga^i\psi_{[a}X^i_{b]},
 \eeq
 and 
 \beq
\de L_{Yuk, no f}|_{\de X}=\tfrac{ig}{32}e(\bar\psi_a\psi_aX_b^k-10\bar\psi_a\psi_bX^k_a+2\bar\psi_a\Ga^{ik}\psi_bX^i_a)(-i\bar\psi_b\Ga^k\ep_m)\,.
\eeq
To find the identity that must be proven by fierzing we drop $\tfrac{eg}{64}X^k_b$ from the above expressions and get (after writing the three-algebra indices in the right way for fierzing)
\beqa
&&2\bar\psi_a\psi_a\bar\psi_b\Ga^k\ep-20\bar\psi_b\psi_a\bar\psi_a\Ga^k\ep-4\bar\psi_b\Ga^{ik}\psi_a\bar\psi_a\Ga^i\ep
+3\bar\psi_a\ga^{\mu}\Ga^{ik}\psi_a\bar\psi_b\ga_{\mu}\Ga^{i}\ep\notag\\[1mm]
&&+\tfrac{1}{2}\bar\psi_a\ga^{\mu}\Ga^{ij}\psi_a\bar\psi_b\ga_{\mu}\Ga^{ijk}\ep
+8\bar\psi_b\ga^{\mu}\psi_a\bar\psi_{a} \ga_{\mu}\Ga^k\ep=0\,.
\eeqa

We now would like to Fierz these terms so that the two fermions with the index $a$ sit together. The possible terms are then
\beq
\bar\psi_a  \psi_a,\,\,\,\bar\psi_a \Ga^{ijkl} \psi_a,\,\,\,\bar\psi_a \ga^{\mu}\Ga^{ij}\psi_a,
\eeq
since the spinors are chiral and only $\Ga^{2n}$ can appear between two spinors of the same chirality. 
Here we need  the symmetrized Fierz identity:
\beq
\psi_{(a}\otimes \psi_{b)}=-\tfrac{1}{16}C\bar\psi_{(a} \psi_{b)}+\tfrac{1}{32}\ga^{\mu}\Ga^{ij}\bar\psi_{(a}\ga_{\mu}\Ga^{ij} \psi_{b)}-
\tfrac{1}{32}\tfrac{1}{4!}C\Ga^{ijkl}\bar\psi_{(a}\Ga^{ijkl} \psi_{b)},
\eeq
which reads for $a=b$
\beq
\psi_{a}\otimes \psi_{a}=-\tfrac{1}{16}C\bar\psi_{a} \psi_{a}+\tfrac{1}{32}\ga^{\mu}\Ga^{ij}\bar\psi_{a}\ga_{\mu}\Ga^{ij} \psi_{a}-
\tfrac{1}{32}\tfrac{1}{4!}C\Ga^{ijkl}\bar\psi_{a}\Ga^{ijkl} \psi_{a}.
\eeq
If we multiply by the 3d charge conjugation matrix from the right and use $\psi C=\bar\psi$ this becomes 
\beq
\psi_{a}\otimes \bar\psi_{a}=-\tfrac{1}{16}\bar\psi_{a} \psi_{a}+\tfrac{1}{32}\ga^{\mu}\Ga^{ij}\bar\psi_{a}\ga_{\mu}\Ga^{ij} \psi_{a}-
\tfrac{1}{32}\tfrac{1}{4!}\Ga^{ijkl}\bar\psi_{a}\Ga^{ijkl} \psi_{a}\,.
\eeq

We now construct the three Fierz expressions needed in the 2nd, 3rd and last terms above: 
\beq
\bar\psi_b|\psi_a\otimes \bar\psi_a|\Ga^k\ep=-\tfrac{1}{16}\bar\psi_b\Ga^k\ep\bar\psi_{a} \psi_{a}+\tfrac{1}{32}\bar\psi_b\ga^{\mu}\Ga^{ij}\Ga^k\ep\bar\psi_{a}\ga_{\mu}\Ga^{ij} \psi_{a}-
\tfrac{1}{32}\tfrac{1}{4!}\bar\psi_b\Ga^{ijkl}\Ga^k\ep\bar\psi_{a}\Ga^{ijkl} \psi_{a}
\eeq
\beqa
&&\bar\psi_b\Ga^{ik}|\psi_a\otimes \bar\psi_a|\Ga^i\ep=\notag\\[1mm]
&&-\tfrac{1}{16}\bar\psi_b\Ga^{ik}\Ga^i\ep\bar\psi_{a} \psi_{a}+\tfrac{1}{32}\bar\psi_b\Ga^{ik}\ga^{\mu}\Ga^{(2)}\Ga^i\ep\bar\psi_{a}\ga_{\mu}\Ga^{(2)} \psi_{a}-
\tfrac{1}{32}\tfrac{1}{4!}\bar\psi_b\Ga^{ik}\Ga^{(4)}\Ga^i\ep\bar\psi_{a}\Ga^{(4)} \psi_{a}~~
\eeqa
and finally
\beqa
&&\bar\psi_b\ga^{\mu}|\psi_a\otimes \bar\psi_a|\ga_{\mu}\Ga^k\ep\notag\\[1mm]
&=&-\tfrac{1}{16}\bar\psi_b\ga^{\mu}\ga_{\mu}\Ga^k\ep\bar\psi_{a} \psi_{a}+\tfrac{1}{32}\bar\psi_b\ga^{\mu}\ga^{\nu}\Ga^{ij}\ga_{\mu}\Ga^k\ep\bar\psi_{a}\ga_{\nu}\Ga^{ij} \psi_{a}-
\tfrac{1}{32}\tfrac{1}{4!}\bar\psi_b\ga^{\mu}\Ga^{ijkl}\ga_{\mu}\Ga^k\ep\bar\psi_{a}\Ga^{ijkl} \psi_{a}\notag\\[1mm]
&=&-\tfrac{3}{16}\bar\psi_b\Ga^k\ep\bar\psi_{a} \psi_{a}-\tfrac{1}{32}\bar\psi_b\ga^{\nu}\Ga^{ij}\Ga^k\ep\bar\psi_{a}\ga_{\nu}\Ga^{ij} \psi_{a}-
\tfrac{3}{32}\tfrac{1}{4!}\bar\psi_b\Ga^{ijkl}\Ga^k\ep\bar\psi_{a}\Ga^{ijkl} \psi_{a}
\eeqa

First we add up the $-\tfrac{1}{32}\tfrac{1}{4!}\bar\psi_b\Ga^{ijkl}\Ga^k\ep\bar\psi_{a}\Ga^{ijkl} \psi_{a}$ terms which cancel since
\beq
-20-4+3\cdot 8=0,
\eeq
and secondly we check the $\bar\psi_a\psi_a\bar\psi_b\Ga^k\ep$ terms which cancel since
\beq
2-20(-\tfrac{1}{16})-4(\tfrac{7}{16})+8(-\tfrac{3}{16})=0,
\eeq
and finally the terms $\bar\psi_b\ga^{\mu}\Ga^{ij}\Ga^k\ep\bar\psi_{a}\ga_{\mu}\Ga^{ij} \psi_{a}$  cancel since
\beq
-20\tfrac{1}{32}\Ga^{ij}\Ga^k-4\tfrac{1}{32}(-4\Ga^k\Ga^{ij}+\Ga^{ij}\Ga^k)+3\Ga^{i}\de^{jk}+\tfrac{1}{2}\Ga^{ijk}+8(-\tfrac{1}{32})\Ga^{ij}\Ga^k=0.
\eeq
Thus we find that the $\ep\psi^3 X$ terms cancel in the supersymmetry variation of the lagrangian.

\section{Closing the SUSY algebra}\label{algebraapp}

\subsection{Conventions}

We essentially use the same conventions as in \cite{Bagger:2007jr}. For example the $\Gamma$-matrices are $11$-dimensional, and using the chirality properties of $\epsilon$, we get
\begin{equation}
\Gamma_{\mu\nu} \epsilon =  \epsilon_{\mu\nu\rho} \Gamma^\rho \epsilon 
\end{equation}
 for dualizing $\Gamma$-matrices with three-dimensional indices (same sign for acting on $\chi_\mu$, opposite sign when acting on $\Psi_a$). For $\Gamma$-matrices with eight-dimensional indices we have the following dualization relation
\begin{equation}
\Gamma_{I_1 \cdots I_p} = \frac{(-1)^{p(p-1)/2}}{(8-p)!} \epsilon_{I_1 \cdots I_{p}}{}^{J_{p+1}\cdots J_{8}}\Gamma_{(8)}~,
\end{equation}
where $\Gamma_{(8)}$ is the product of all $\Gamma$-matrices with eight-dimensional indices.

\subsection{$[ \delta_1,\delta_2] \Psi_a$}

Here we will give some more details on closing the SUSY algebra on $\Psi_a$ focussing on one of the main differences compared to the analysis in  \cite{Bagger:2007jr}, namely that we need to allow for superconformal transformations in the right hand side of the algebra. Under superconformal transformations $\Psi_a$ transforms as
\begin{equation}
\delta_S \Psi_a = X^I_a \Gamma^I \eta
\end{equation}
where
\begin{eqnarray}
\eta &=& \tfrac{i}{32} \Big(  \gamma_1 (\bar\epsilon_1 \Gamma_\mu \epsilon_2) \Gamma^\mu \Gamma^J X^J_b \Psi_b +\tfrac{\gamma_2}{2!} (\bar\epsilon_1 \Gamma_{LM} \epsilon_2) \Gamma^{L}  X^M_b \Psi_b \notag \\
&&+\tfrac{\gamma_3}{2!} (\bar\epsilon_1 \Gamma_{LM} \epsilon_2) \Gamma^{LMN}  X^N_b \Psi_b +\tfrac{\gamma_4}{4!} (\bar\epsilon_1 \Gamma_\mu\Gamma_{L_1 \cdots L_4} \epsilon_2) \Gamma^\mu\Gamma^{L_1 L_2 L_3}  X^{L_4}_b \Psi_b \\
&&+\tfrac{\gamma_5}{4!} (\bar\epsilon_1 \Gamma_\mu\Gamma_{L_1 \cdots L_4} \epsilon_2) \Gamma^\mu\Gamma^{L_1 \cdots L_4 M}  X^{M}_b \Psi_b  \Big) \notag ~.
\end{eqnarray}

We first analyse the $\Gamma^{(5)}$ term in (\ref{ddpsi}). As can be seen from (\ref{parameters}) this term does not represent symmetries of the theory and it must therefore vanish. As a consistency check we note that all terms containing the three-algebra structure constants cancel against each other, just as in the BLG analysis \cite{Bagger:2007jr}. There are three different gamma-structures represented by the ellipses in (\ref{ddpsi}). The ones with two and six $SO(8)$ gamma matrices, which contain just one type of term each, combine via duality to give
\begin{eqnarray}
\alpha_3-2\alpha_4+4\alpha_5-2\alpha_6-\alpha_7 -\tfrac{1}{8}\gamma_4-\tfrac{1}{2}\gamma_5= 0~.\label{G55solution}
\end{eqnarray}
The terms coming with four $SO(8)$ gamma matrices contain two independent structures with respect to the three-algebra indices. One of these structures further split into three independent terms which, however, give rise to the same equation. In this way we get
\begin{eqnarray}
-2\alpha_6 + \alpha_7 &=&0~,\notag\\
\alpha_3+2\alpha_4-4\alpha_5-2\alpha_6-\alpha_7 -\tfrac{1}{8}\gamma_4+\tfrac{1}{2}\gamma_5&=&0~.
\end{eqnarray} 
We can now combine the equations and summarize the constraints from the $\Gamma^{(5)}$ term in (\ref{ddpsi}) as
\begin{eqnarray}
\alpha_3 -2 \alpha_7 &=&0~,\notag\\
-\alpha_4+2\alpha_5 -\tfrac{1}{16}\gamma_4-\tfrac{1}{4}\gamma_5&=& 0~,\\
-2\alpha_6 + \alpha_7 &=&0~.\notag
\end{eqnarray}

We now turn to the $\Gamma^{(1)}$ term in (\ref{ddpsi}). From (\ref{parameters}) it follows that the only contributing symmetry transformation is the transport term, and the reminder becomes the Dirac equation
\begin{eqnarray}
&&0=\slashed{\tilde D}\Psi_a +\tfrac{1}{2}\Gamma_{IJ}X^I_c X^J_d f^{cdb}{}_a \Psi_b +(\tfrac{\alpha_2}{\alpha_1}+\alpha_5)X^2 \Psi_a\notag \\
&&~~~~~~~+({2\alpha_5+\tfrac{3}{2}\alpha_7 -\tfrac{1}{32}\gamma_1-\tfrac{7}{128}\gamma_4-\tfrac{7}{32}\gamma_5})X^I_a X^I_b \Psi_b  \\
&&~~~~~~~+(2\alpha_5-\tfrac{1}{32}\gamma_1-\tfrac{7}{128}\gamma_4-\tfrac{7}{32}\gamma_5)\Gamma_{IJ}X^I_a X^J_b \Psi_b - \tfrac{1}{\alpha_1} X^I_a \Gamma^I \Gamma^\mu f_\mu ~.\notag
\end{eqnarray}

The remaining part of  (\ref{ddpsi}) is the $\Gamma^{(2)}$ term, which now turns into a consistency condition as it has to be written in terms of the Dirac equation above \cite{Bagger:2007jr}, and the remaining symmetry parameters in  (\ref{parameters}). This leads to six equations, one coming from a structure with four gamma matrices, and five coming from structures containing two gamma matrices. The condition from the structure with four gamma matrices yields
\begin{eqnarray}
\alpha_5 +\tfrac{1}{4}\alpha_7 +\tfrac{1}{64}\gamma_1+\tfrac{1}{32}\gamma_3-\tfrac{11}{256}\gamma_4-\tfrac{11}{64}\gamma_5=0~.\label{G4constr}
\end{eqnarray}
The equations coming from the linearly independent structures with two gamma matrices are
\begin{eqnarray}
\alpha_5 +\tfrac{1}{4}\alpha_7 +\tfrac{1}{64}\gamma_1+\tfrac{1}{32}\gamma_3-\tfrac{11}{256}\gamma_4-\tfrac{11}{64}\gamma_5&=&0~,\notag \\\label{G2constr}
\alpha_5-\tfrac{1}{4}\alpha_7 +\tfrac{1}{64}\gamma_1+\tfrac{1}{64}\gamma_2-\tfrac{11}{256}\gamma_4-\tfrac{11}{64}\gamma_5&=&0~,\notag\\
-3\alpha_5-\tfrac{1}{4}\alpha_7-\tfrac{1}{64}\gamma_1+\tfrac{1}{32}\gamma_3+\tfrac{23}{256}\gamma_4+\tfrac{23}{64}\gamma_5&=&0~,\\
4\tfrac{\alpha_2}{\alpha_1}-8\alpha_5-\alpha_7&=&0~,\notag\\
\alpha_7-2\alpha_6&=&0~.\notag
\end{eqnarray}
Solving all the above equations yield
\begin{eqnarray}
\tfrac{\alpha_2}{\alpha_1} &=& 2 \alpha_5+ \tfrac{1}{4}\alpha_7 \notag ~,\\
\alpha_3 &=& 2 \alpha_7 \notag ~,\\
\alpha_6 &=& \tfrac{1}{2}\alpha_7 \notag~, \\
\gamma_1 &=& -68 \alpha_4 + 8 \alpha_5 -16 \alpha_7  \notag ~,\\
\gamma_2 &=&  24 \alpha_4 + 16 \alpha_5 + 32 \alpha_7 \notag ~,\\
\gamma_3 &=&  12 \alpha_4 + 8 \alpha_5 \notag ~,\\
\gamma_4 &=& -16\alpha_4 +32 \alpha_5 -4 \gamma_5 \notag ~.\\
\end{eqnarray}

\subsection{Summary}

The resulting Dirac equation becomes
\begin{eqnarray}
&&0=\slashed{\tilde D}\Psi_a +\tfrac{1}{2}\Gamma_{IJ}X^I_c X^J_d f^{cdb}{}_a \Psi_b -\tfrac{\alpha}{8}X^2 \Psi_a +\tfrac{5}{4}\alpha X^I_a X^I_b \Psi_b \notag \\
&&~~~~~~~ -\tfrac{1}{4}\alpha\Gamma_{IJ}X^I_a X^J_b \Psi_b - \tfrac{1}{\alpha_1} X^I_a \Gamma^I \Gamma^\mu f_\mu ~.
\end{eqnarray}

and the supersymmetry variations, after closing the algebra on all the fields, takes the form
\begin{eqnarray}
\delta e_{\mu}{}^{\alpha}&=& \alpha_1 i\bar\epsilon \Gamma^{\alpha}\chi_{\mu}~, \notag\\
\delta\chi_{\mu}&=&\alpha_1\tilde D_{\mu}\epsilon + {\cal O}(\chi^2)~,\notag\\
\delta B_{\mu}^{IJ}&=&- \frac{i}{2}\alpha_1 e^{-1}\bar\epsilon\Gamma^{IJ}\Gamma_{\nu}\Gamma_{\mu}f^{\nu}+2 i \alpha\bar \Psi_a \Gamma_\mu \Gamma^{[I}\epsilon X^{J]}_a-\tfrac{i}{4}  \alpha\bar \Psi_a \Gamma^K\Gamma^{IJ}\Gamma_\mu \epsilon X^K_a \notag\\
&&+ 2i \tfrac{\alpha}{\alpha_1}  (\bar\chi_\mu \Gamma^{K[I}\epsilon) X^{J]}_a X^K_a+ \tfrac{i}{4} \tfrac{\alpha}{\alpha_1}(\bar\chi_\mu \Gamma^{IJ} \epsilon) X^2 +{\cal O}(\chi^2),\notag\\
\delta X_{I}^a&=& i \bar \epsilon \Gamma^I \Psi_a~,\notag\\[1mm]
\delta\Psi_a&=&-(\tilde D_{\mu}X^I_a-\frac{i}{\alpha_1}\bar \chi_\mu \Gamma^I \psi_a )\Gamma^{\mu}\Gamma^I \epsilon+\tfrac{1}{6}X^I_b X^J_c X^K_d \Gamma^{IJK}\epsilon f^{bcd}{}_a\\&&~-\tfrac{1}{8}\alpha \Gamma^I\epsilon X^I_a X^2+\tfrac{1}{2}\alpha \Gamma^I \epsilon X^J_a X^J_b X^I_b+ {\cal O}(\chi^2)~,\notag\\[1mm] 
\delta \tilde A_{\mu}{}^b{}_a &=&  i  \bar\epsilon \Gamma_{\mu} \Gamma^I X_c^I \Psi_d f^{cdb}{}_a- i \alpha  \bar \epsilon \Gamma_\mu \Gamma^IX^I_{[b}\Psi_{a]}+  \tfrac{i}{\alpha_1} (\bar\epsilon \Gamma^{IJ}\chi_\mu) X^I_c X^J_d f^{cdb}{}_a\notag\\
&&-i\tfrac{\alpha}{\alpha_1} (\bar\epsilon \Gamma^{IJ}\chi_\mu) X^I_{b}X^J_a+ {\cal O}(\chi^2)~,\notag
\end{eqnarray}
where $\alpha_1=\pm \sqrt{2}$ and we have replaced $\alpha_7$ by just $\alpha$.

\section{Superconformal geometry}

\subsection{Conformal constraints}


$\cN$-extended conformal supergravity can be described in terms of the geometry of a supermanifold $M$ with (even$|$odd)-dimension $(3|2\cN)$. As is usual in superspace, it is convenient to work in a preferred basis specified by a set of basis forms $E^{\cA}=(E^a,E^{\a I})$, where $a=0,1,2$ is a Lorentz vector index, $\a=1,2$ is a spinor index and $I=1,\ldots \cN$ is an $SO(\cN)$ vector index, and their vector duals. The structure group is $SL(2,\bbR)\xz SO(\cN)$, and the connection $\O_\cA{}^\cB$ takes its values in the Lie algebra of this group.
The torsion and curvature are defined as usual.\footnote{The conventions used in the superspace sections can be found in  \cite{Greitz:2011vh}, for general $\cN$ and \cite{Greitz:2012vp} for $\cN=8$, although $i,j$ are used there for $SO(\cN)$ vector indices instead of $I,J$.}The only constraint necessary to describe off-shell superconformal geometry is

\beq
 T_{\a I\b J}{}^c=-i\d_{IJ} (\c^c)_{\a\b}\,;\qquad c=0,1,2\ .
 \label{2.1}
\eeq

Using the Bianchi identities and all possible conventional constraints, which correspond to choosing the connection and the even vector basis $E_a$, we find that the only other non-zero torsion components are

\beqa
 T_{a \b J}{}^{\c K}&=&   (\c_a)_\b{}^\c K_J{}^K + (\c^b)_\b{}^\c L_{ab J}{}^K\ ,
 \label{2.6}
\eeqa
where $K_{IJ}$ is symmetric and $L_{abIJ}$ is antisymmetric on both pairs of indices, and the dimension three-halves torsion whose leading component is the gravitino field strength. The dimension-one curvatures are

\beqa
 R_{\a I\b J, cd}&=& -2i(\c_{cd})_{\a\b} K_{IJ} -2i\ve_{\a\b} L_{cdIJ}\nn\w1
 R_{\a I\b J,KL}&=& i\ve_{\a\b}(M_{IJKL} + 4 \d_{[I[K} K_{J]L]})-i(\c^a)_{\a\b} (4\d_{(I[K} L_{a J) L]}-\d_{IJ} L_{a KL})\ ,
 \label{2.7}
\eeqa
where $L_{ab}=\ve_{abc} L^c$, and $M_{IJKL}$ is totally antisymmetric. The dimension three-halves Lorentz curvature is

\beq
R_{a\b J,cd}=-\frac{i}{2}(\c_a\Psi_{cd}-2\c_{[c} \Psi_{d]a})_{\b J}\ ,
\label{2.8}
\eeq
where the dimension three-halves torsion has been rewritten as $\Psi_{ab\c K}$. The $SO(\cN)$ curvature, $R_{a\b J,KL}$ has gamma-traceless and spinor parts given by

\beqa
\hat R_{a\b I,JK}&=& \chi_{a\b IJK}-i\d_{I[J} \hat\Psi_{a\b K]}\nn\w1
R_{\a I,JK}&=&\r_{\a I,JK}-2\l_{\a IJK}+2\d_{I[J} \r_{\a K]}\ ,
\label{2.9}
\eeqa
where we have decomposed the dual of the gravitino field strength as $\Psi_a=\hat\Psi_a + \c_a \Psi$ and where $\r_{I,JK}$ and $\l_{IJK}$ are in the irreducible (i.e. traceless) tableaux $\Yvcentermath1\tiny\yng(2,1)$ and $\Yvcentermath1\tiny\yng(1,1,1)$. The field $\chi$ is also totally antisymmetric as well as being Lorentz gamma-traceless. The derivative of $K_{IJ}$ is given by

\beq
D_{\a I} K_{JK}=2\r_{\a (J,K)I}+ 2\d_{I(J} \k_{\a K)} + \d_{JK} \k'_{\a I}\  ,
\label{2.10}
\eeq
while the derivative of $L_{a IJ}$ is

\beq
D_{\a I} L_{aJK}=\chi_{a\a IJK}+i\d_{I[J}\hat\Psi_{a\a K]} +(\c_a)_\a{}^\b(\l_{IJK} + \r_{I,JK}+2 \d_{I[J} \s_{K]})_\b\ ,
\label{2.11}
\eeq
The spin one-half fields in the vector representation of $SO(8)$ are related by

\beq
\k=\frac{i}{2}\Psi\ ,\qquad \k'=2\s-\frac{i}{4}\Psi\ ,\qquad \r=\s+\frac{i}{2}\Psi\ .
\label{2.12}
\eeq
In addition, we have

\beq 
D_{\a I} M_{JKLM}=i\l_{IJKLM} + 12i\d_{I[J} \l_{\a KLM]}\ .
\label{2.13}
\eeq

This geometry  describes an off-shell superconformal multiplet \cite{Howe:1995zm}. The interpretation of the dimension-one fields, $K,L,M$, is as follows. The geometry is determined by the basic constraint \eq{2.1} which is invariant under Weyl rescalings where the parameter is an unconstrained scalar superfield. This means that some of the fields that appear in the geometry do not belong to the conformal supergravity multiplet. At dimension one $K$ and $L$ are of this type, so that we could set them to zero if we were only interested in the superconformal multiplet. The field $M_{IJKL}$, on the other hand, can be considered as the field strength superfield for the conformal supergravity multiplet \cite{Howe:1995zm}. Similarly, at dimension three-halves, the fields $\l_{IJK}$ and $\l_{IJKLM}$ are components of the Cotton superfield, while $\s,\r$ and $\chi$ are like $K$ and $L$ in that their leading components can be removed by super-Weyl transformations \cite{Howe:1995zm,Howe:2004ib,Kuzenko:2011xg}. It is easy to see that these fields correspond to the $\th^3$ components of a scalar superfield.

The fact that $M$ is not expressible in terms of  the torsion is due to a lacuna in Dragon's theorem  \cite{Kuzenko:2011xg,Cederwall:2011pu} which in higher-dimensional spacetimes states that the curvature is so determined \cite{Dragon:1978nf}. We recall that in three-dimensional spacetime there is no Weyl tensor but that its place is taken by the dimension-three Cotton tensor. This turns out to be a component of the field $M_{IJKL}$ so that we could refer to the latter as the super Cotton tensor. Using the notation $[k,l]$ to denote fields that have $k$ antisymmetrised $SO(\cN)$ indices and $l$ symmetrised spinor indices, one can see that the component fields of the superconformal multiplet fall into two sequences starting from $M_{IJKL}$ \cite{Greitz:2011vh}. The first has fields of the type $[4-p,p]$, where the top ($[4,0]$) component is the supersymmetric Cotton tensor, while the second has fields of the type $(4+p,p)$ and therefore includes higher spin fields for $\cN>8$. There is also a second scalar $[4,0]$ at dimension two. Fields with two or more spinor indices obey covariant conservation conditions so that each field in the multiplet has two degrees of freedom multiplied by the dimension of the $SO(\cN)$ representation, provided that we count the dimension one and two scalars together. It is easy to see that the number of bosonic and fermionic degrees of freedom in this multiplet match. Diagrammatically, we have the following picture:

\vskip 1cm

\begin{picture}(400,200)
\put(202,180){[4,0]}
\put(210,175){\vector(-1,-1){20}}\put(215,175){\vector(1,-1){20}}
\put(175,145){[5,1]}\put(235,145){[3,1]}
\put(170,135){\vector(-1,-1){20}}\put(180,135){\vector(1,-1){20}}\put(245,135){\vector(-1,-1){20}}\put(255,135){\vector(1,-1){20}}
\put(135,100){[6,2]}\put(202,100){[4,0]}\put(275,100){[2,2]}
\put(130,90){\vector(-1,-1){20}}\put(295,90){\vector(1,-1){20}}
\put(95,55){[7,3]}\put(315,55){[1,3]}
\put(85,45){\vector(-1,-1){20}}\put(335,45){\vector(1,-1){20}}
\put(355,10){[0,4]}
\end{picture}

\begin{center}Figure 1\end{center}

\vskip 1cm

The dimension of the top field (i.e. $M_{IJKL}$)  is one and thereafter the dimension increases stepwise by one-half as one goes down the diagram. The spins of the fields are given by the second entry divided by two. For $\cN<4$ the top field will be the one with $\cN$ internal indices; for example, in $\cN=3$ it will be the dimension three-halves field $[3,1]$. The right sequence clearly terminates at $[0,4]$ but the left sequences can continue to higher spin for $\cN>8$. The fields $[2,2],[3,1]$ and $[0,4]$ are the $SO(\cN)$ gauge field strength, the supersymmetric partner of the Cotton tensor (Cottino), and the Cotton tensor respectively. In the case of $\cN=6$ there is an additional $U(1)$ fields strength $[6,2]$ that plays a key 
r\^ole in the ABJM model.  It is therefore permissible in this case to introduce a new field strength two-form $G$ that satisfies an abelian Bianchi, $dG=0$. At dimension one we can take

\beq
G_{\a I\b J}=i\ve_{\a\b} M_{IJ}\ ,
\label{2.15}
\eeq
where $M_{IJ}:=\frac{1}{4!}\ve_{IJKLMN} M^{KLMN}$ is the dual of the four-index scalar appearing in the dimension-one $SO(6)$ curvature. The dimension three-halves Bianchi identity for $G$ then implies that

\beq
D_{\a I} M_{JK}=2i \d_{I[J} \l_{\a K]} + 3i\tilde \l_{\a IJK}\ ,
\label{2.16}
\eeq
where $\tilde\l_{IJK}$ is the dual of $\l_{IJK}$ and $\l_I$ is the dual of $\l_{IJKLM}$. Indeed, \eq{2.16} is equivalent to \eq{2.13} for $\cN=6$. The dimension three-halves component of $G$ is

\beq
G_{a\b J}=-i(\c_a\l_J)_\b\ .
\label{2.17}
\eeq

In $\cN=8$ one can impose a duality condition on the dimension-one scalar fields that halves the multiplet; the fields in the left sequence become the duals of those in the right sequence. The dimension-two scalar fields also obey a duality constraint but it is opposite to that for the dimension-one scalars. A consequence of this is that there is no off-shell Lagrangian for the $\cN=8$ theory. 

In addition, in $\cN=8$ we can take the R-symmetry group to be $Spin(8)$ rather than $SO(8)$ \cite{Cederwall:2011pu,Greitz:2012vp}. It turns out that this is the correct choice in order to describe the theories we are interested in, and so we shall switch to this for the remainder of the paper. We denote the spinor indices by $A,B,\ldots$ ($(0010)$ representation) and $A'.B'\ldots$ ($(0001)$ representation), while we keep $I,J,\ldots$ for the vector representation $(1000)$. All three types of index can take 8 values.  So for $\cN=8$ we shall take the basis odd one-forms to be $E^{\a A}$, and in the above formulae for the components of the torsion and curvature tensors replace all the internal vector indices by unprimed spinorial ones. 

The super Cotton tensor $M_{ABCD}$ will be chosen to be self-dual, i.e. in the representation $(2000)$, which is equivalent to a symmetric traceless second-rank tensor denoted by $C_{IJ}$:

\beq
M_{ABCD}=\frac{1}{16}(\S^{IK})_{[AB} (\S^J{}_K)_{CD]} C_{IJ}\ ,
\label{2.21}
\eeq
or, inverting,

\beq
C_{IJ}=\frac{1}{24}(\S_{IK})_{AB} (\S_J{}^K)_{CD} M^{ABCD}\ .
\label{2.22}
\eeq

Furthermore, we can relate the algebra indices on the curvature in three eight-dimensional representations by means of $\S$-matrices:

\beq
R_{AB}=\frac{1}{4}(\S^{IJ})_{AB} R_{IJ}\, ;\qquad  R_{A'B'}=\frac{1}{4}(\S^{IJ})_{A'B'} R_{IJ}\ .
\label{2.221}
\eeq
Inverting, we have

\beq
R_{IJ}=\frac{1}{4}(\S_{IJ})^{AB} R_{AB}\, =\frac{1}{4}(\S_{IJ})^{A'B'} R_{A'B'}\ .
\label{2.222}
\eeq
Clearly similar formulae apply to any second-rank anti-symmetric tensors, such as $L_{a IJ}$.

The dimension three-halves curvatures and relations are given by equations \eq{2.9} to \eq{2.13} but with $I,J,K,..$ replaced by $A,B,C$. The five-index $\l$ spinor is the dual of $\l_{ABC}$ multiplied by a factor of $1/3$. 	The field $\l_{ABC}$ can also be written as a $\S$-traceless primed vector-spinor $\l_{IA'}$, with

\beq
\l_{IA'}=-\frac{1}{3}(\S^J)_{A'}{}^A(\S_{IJ})^{BC}\l_{ABC}\ ,
\label{2.25}
\eeq
and it is easy to check that

\beq
\l_{\a IA'}=\frac{i}{15}(\S^J)_{A'}{}^A D_{\a A} C_{IJ}\ .
\label{2.26}
\eeq

For the Yang-Mills sector there is a similar off-shell multiplet which is relevant to the superconformal case, i.e. the Chern-Simons Lagrangian. The $(0,2)$ component of the field strength two-form $F$ is taken to be \cite{Samtleben:2009ts}

\beq
F_{\a I\b J}=i\ve_{\a\b} W_{IJ}\ ,
\label{2.261}
\eeq
where $W_{IJ}$, which is in the adjoint representation of the gauge group, is antisymmetric on its $SO(\cN)$ indices (and we assume $\cN\geq2$). The lowest-order Bianchi identity will be satisfied if

\beq
D_{\a I} W_{JK}=D_{\a [I} W_{JK]} -\frac{2}{\cN-1} \d_{I[J} D_\a^L W_{K]L}\ ,
\label{2.262}
\eeq
where the derivative is now covariant with respect to the gauge group as well as the geometry. The multiplet described by this constraint is rather similar to the super Cotton multiplet. Its independent components can be represented by a similar diagram with top vertex given by a field $[2,0]$ which is the leading component of $W_{IJ}$:

\vskip 1cm

\begin{picture}(400,90)
\put(202,90){[2,0]}
\put(210,85){\vector(-1,-1){20}}\put(215,85){\vector(1,-1){20}}
\put(175,55){[3,1]}\put(235,55){[1,1]}
\put(170,50){\vector(-1,-1){20}}\put(180,50){\vector(1,-1){20}}\put(245,50){\vector(-1,-1){20}}\put(255,50){\vector(1,-1){20}}
\put(135,20){[4,2]}\put(202,20){[2,0]}\put(275,20){[0,2]}
\put(130,15){\vector(-1,-1){20}}
\end{picture}\begin{center}Figure 2\end{center}

The fields on the right diagonal terminate at $[0,2]$, i.e. $F_{ab}$, while the fields on the left diagonal generically involve higher-spin components that obey covariant divergence constraints. This multiplet is off-shell but can only be used to construct an off-shell Lagrangian for the cases $\cN=2,3,4$ (with self-duality imposed for the latter), but not for the cases we are interested in. (Although there is an off-shell version of Chern-Simons gauge theory in $\cN=6$ harmonic superspace \cite{Howe:1994ms}; for other discussions of Chern-Simons theories in harmonic superspaces, including ABJM in $\cN=3$ harmonic superspace, see e.g. \cite{Zupnik:2007he,Buchbinder:2008vi}.)

\subsection{Completeness of the solution}


It is easy enough to see at the linearised level that the components in the Cotton superfield with spins $\geq 1$, i.e. two or more symmetrised spinor indices, must also obey conservation conditions which must become covariant conservation constraints  in the full theory because otherwise the degrees of freedom account would go astray. This has not been verified directly, however, except for the Cottino and Cotton tensors in the case of $\cN=8$. What can be said, however, is that there are no more constraints coming from higher-dimensional Bianchi identities, and that we therefore have a complete solution to these equations, even for $\cN>8$. 

We can show this making use of the idea  which is based on the fact that the Bianchi identities themselves obey identities even when they are not satisfied \cite{Sohnius:1980iw}.\footnote{We assume here that $\cN>4$ although only minor modifications are needed for the other cases.} These are

\beq
D\cI=0:\qquad D\cI^\cA=-E^\cB \cI_\cB{}^\cA;\qquad D\cI_\cA{}^\cB=0\ ,
\label{2.14}
\eeq
where $\cI\, ,\cI^\cA$ and $\cI_\cA{}^\cB$ are respectively the Bianchi identities for the Yang-Mills field strength, the torsion and the curvature, and
where we have made use of the Ricci identity, $D^2\sim F+R$. The idea now is to show that when a subset of the Bianchi identities have been satisfied then the other, higher-dimensional ones automatically are by virtue of \eq{2.14}. For this we shall need to use a little bit of superspace cohomology. The basic idea is that a given $n$-form can be split into $(p,n-p)$ bi-degrees where the pair $(p,q)$ denotes the number of even (odd) indices on a particular component, with increasing $p$ corresponding to increasing mass dimension. Moreover, the exterior derivative $d$ splits into components with bi-degrees $(-1,2)$, $(0,1)$, $(1,0)$ and $(2,-1)$. The first of these, denoted $t_0$, has dimension zero and is purely algebraic. On a $(p,q)$ form it operates by contracting one of the even indices with the vector index on the dimension-zero torsion and then by symmetrising over the $q+2$ odd indices thereby giving a form of bi-degree $(p-1,q+2)$. It squares to zero and is thus associated with its own cohomology $H_t^{p.q}$ \cite{Bonora:1986ix}. It is the only one we shall need in the following. Suppose we have an equation of the form $d\o=0$ where $\o$ is an $n$-form, and suppose that the lowest non-vanishing component of $\o$ is $\o_{p,q}$. Then the lowest non-trivial component of $d\o=0$ is $t_0 \o_{p,q}=0$, and is thus purely algebraic. We shall always assume that the dimension-zero torsion is as given in \eq{2.1} and so solving this equation is the same as in flat superspace. In three dimensions, for $\cN>2$, the key fact is that $H_t^{p,q}=0$ for $p>0$. This can be seen via dimensional reduction \cite{Berkovits:2008qw,Movshev:2011pr} or directly in three dimensions \cite{Brandt:2010fa}.

A simple example is given by the Yang-Mills case. Suppose that we have solved the lowest-dimensional component, i.e. $\cI_{0,3}=0$. Then $D\cI=0$ implies that $t_0\cI_{1,2}=0$. Since $H_t^{1,2}=0$, we must have $I_{1,2}=t_0 J_{2,0}$, so that solving this Bianchi identity component by setting $J_{2,0}=0$ simply allows one to solve for $F_{2,0}$ in terms of the lower-dimensional fields. Then with $\cI_{1,2}=0$ similar arguments show that $\cI_{2,1}=\cI_{3,0}=0$, in other words, a complete solution to the identities is guaranteed if $\cI_{0,3}=0$, i.e. if \eq{2.262} holds.

For the geometry, we shall suppose that the torsion Bianchi identities have been solved up to dimension three-halves, so that the torsion and dimension-one curvatures have the form given above. (The dimensions of the $(p,q)$ components of the identities are given by $(p+q/2)-k$ where $k=1,1/2,0$ for $\cI^a, \cI^{\ua}, \cI_\cA{}^\cB$ respectively, where $\ua$ denotes a combined spinor-internal index.) Furthermore, the dimension three-halves field strengths are determined in terms of derivatives of the dimension-one fields. Using the second of equations \eq{2.14} one can show that the dimension three-halves identity $\cI_{0,3}{}^{ab}$ must be identically satisfied, although this is not true for $\cI_{0,3}{}^{IJ}$. The remaining condition that comes from this identity component is the constraint on the derivative of the super Cotton tensor 
\eq{2.13}. We can now use the third equation in \eq{2.14} to show that the dimension-two components of $\cI_\cA{}^\cB$, namely $\cI_{1,2}{}^{ab}$ and $\cI_{1,2}{}^{IJ}$, are solved by specifying the $(2,0)$ Lorentz and $SO(\cN)$ curvature components in terms of derivatives and bilinears in the dimension-one fields, as in the Yang-Mills case. The second of equations \eq{2.14} can then be used to show that the torsion identities are identically satisfied at dimension-two, and it is a simple matter to confirm that there are no new conditions arising at dimension five-halves, again with the aid of \eq{2.14}. 

The upshot of this analysis is that all of the non-zero components of the torsion and curvature tensors are determined as functions of the independent dimension-one fields $K,L$ and $M$ and their derivatives, and that the super Cotton tensor satisfies the constraint \eq{2.13}.

\subsection{Super-Weyl covariance}


The modified BLG theory given above couples the superconformal geometry to the $\cN=8$ matter system which is superconformal in the flat space limit. We therefore expect the combined system to be invariant under super-Weyl transformations as well as super-diffeomorphisms, but this is not manifest in the formalism. In this subsection we show that the Dirac equation \eq{4.8} is indeed super-Weyl covariant in that it scales homogeneously under super-Weyl transformations.

The latter can be defined as follows. For any variation of the supervielbein and connections we define

\beqa
H_\cA{}^\cB&=&E_\cA{}^\cM \d E_\cM{}^\cB\nn\w1
\F_\cA&=& E_\cA{}^\cM \d \O_\cM\ ,
\label{4.13}
\eeqa
where $\O_M$ is either the Lorentz or $SO(8)$ connection.  The variation of the torsion is given by

\beq
\d T_{\cA\cB}{}^{\cC}=2D_{[\cA} H_{\cB]}{}^{\cC}- T_{\cA\cB}{}^{\cD} H_{\cD}{}^{\cC}-2H_{[\cA}{}^{\cD} T_{|\cD|\cB]}{}^{\cC} + 2\F_{[\cA,\cB]}{}^{\cC}\ ,
\label{4.14}
\eeq
where the antisymmetrisation brackets are graded. In order to preserve the constraints on the torsion under a super-Weyl transformation with parameter an unconstrained superfield $S$ we must take

\beqa
H_{\a A}{}^{\b B}&=&-\d_\a{}^\b \d_A{}^B S \qquad ; \qquad H_a{}^b= -2 \d_a{}^b S\nn\w1
H_a{}^{\b B}&=& -2i(\c_a)^{\b\c} D_\c^B S\ ,
\label{4.15}
\eeqa
and, for the connections,

\beqa
\F_{\a A,bc}&=& 2(\c_{bc})_\a{}^\b D_{\b A} S\nn\w1
\F_{\a A,BC}&=&-4\d_{A[B} D_{\a C]} S\nn\w1
\F_{a,bc}&=&4 \h_{a[b} D_{c]} S\nn\w1
\F_{a,BC}&=&2i D^2_{aBC} S\ ,
\label{4.16}
\eeqa
where

\beq
D_{\a A} D_{\b B} S=\frac{i}{2} \d_{AB} (\c^a)_{\a\b} D_a S + \ve_{\a\b} D^2_{AB} S + (\c^a)_{\a\b} D^2_{a AB} S
\label{4.17}
\eeq
defines the second-order derivatives which are respectively symmetric and anti-symmetric on their $SO(8)$ indices. The dimension-one supergravity fields transform as

\beqa
\d C_{IJ}&=& 2 S C_{IJ}\nn\w1
\d K_{AB}&=& 2 S K_{AB}- 2i D^2_{AB} S\nn\w1
\d L_{a BC}&=& 2S L_{a BC}- 2i D^2_{a BC} S\ .
\label{4.18}
\eeqa

The last two equations here show that the leading components of $K$ and $L$ can be transformed away using the $\th^2$ components of $S$. 
For the matter fields we have

\beqa
\d \f_I&=& S \f_I\nn\w1
\d \L_{\a A'}&=& 2S \L_{\a A'} -i( (\S^I)_{AA'} D_{\a A} S)\f_I\  ,
\label{4.19}
\eeqa
and we take the one-form gauge potentials to be invariant, so that the components with respect to a preferred basis transform only because of the supervielbein factors. In order to compute the super-Weyl variation of the Dirac equation we also need the variation of the gravitino field strength $\Psi$ We have

\beq
T_{ab}{}^{\c  C}:=\ve_{abc}\Psi^{c \c C}
\label{4.20}
\eeq
and we set 

\beq
\Psi_a:=:\hat\Psi _a + \c_a\Psi
\label{4.21}
\eeq
in terms of irreducible Lorentz representations. From the variation of the dimension three-halves torsion we find

\beq
\d\Psi_{\a A}=3 S\Psi _{\a A}+\frac{4i}{3} (\c^a D_a DS)_{\a A}- 4i K_{AB}D_{\a B} S\ .
\label{4.22}
\eeq
If one now simply varies all of the terms in the Dirac equation using the above formulae one finds, after a few pages of algebra, that the whole equation transforms with a factor of $3 S$, reflecting the fact that it is a dimension three-halves equation.

\end{document}